\documentclass[preprintnumbers,floatfix,prd,nofootinbib,twocolumn]{revtex4} 

\usepackage{graphicx}
\usepackage{latexsym}
\usepackage{amssymb}
\usepackage{stmaryrd}
\usepackage{soul}

\usepackage{amsmath}

\DeclareMathOperator{\arcsinh}{arcsinh}

\def\beq{\begin{equation}}
\def\eeq{\end{equation}}
\def\e{\varepsilon}

\def\pmb#1{\setbox0=\hbox{$#1$}%
  \kern-.025em\copy0\kern-\wd0
  \kern.05em\copy0\kern-\wd0
  \kern-.025em\raise.0433em\box0}

\begin{document}

\title{Gravitational scattering of two black holes\\ at the fourth post-Newtonian approximation}

\author{Donato \surname{Bini}$^1$}
\author{Thibault \surname{Damour}$^2$}
 
\affiliation{$^1$Istituto per le Applicazioni del Calcolo ``M. Picone'', CNR, I-00185 Rome, Italy\\
$^2$Institut des Hautes Etudes Scientifiques, 91440 Bures-sur-Yvette, France}

\date{\today}

\begin{abstract}
We compute the (center-of-mass frame) scattering angle $\chi$ of  hyperboliclike  encounters of two spinning black holes, at the fourth post-Newtonian  approximation level for orbital effects, and at the next-to-next-to-leading order for spin-dependent effects. 
We find it convenient to compute the gauge-invariant scattering angle (expressed as a function of energy, orbital angular momentum and spins) by using the Effective-One-Body formalism. The contribution to  scattering associated with nonlocal, tail effects is computed by generalizing
to the case of unbound motions the method of time-localization of the action introduced in the case of (small-eccentricity) bound motions by Damour, Jaranowski
and Sch\"afer [Phys.\ Rev.\ D {\bf 91}, no. 8, 084024 (2015)].
\end{abstract}

\maketitle

\section{Introduction}
The recent observation of gravitational wave signals emitted by the merger of binary black holes \cite{Abbott:2016blz,Abbott:2016nmj,Abbott:2017vtc} makes it very timely to further improve 
the analytical understanding of the gravitational interaction of  compact binary systems in General Relativity (GR). Indeed, 
the construction~\cite{Taracchini:2013rva,Bohe:2016gbl},
within the Effective-One-Body (EOB) formalism~\cite{Buonanno:1998gg,Buonanno:2000ef,Damour:2000we,Damour:2001tu} of 
a large bank of (semi-)analytical binary black-hole merger templates has been crucial in the search, significance assessment and parameter estimation of the merger signals.

The motion of comparable-mass binary systems has been tackled by several different approximation methods: i) post-Newtonian (PN) calculations, ii) post-Minkowskian (PM) calculations and iii) numerical relativity (NR) simulations.
In addition, the EOB formalism is a framework within which the results of all the above approximation methods can be usefully combined, thereby extending the realm of validity of the original methods.
The traditional way in which the EOB formalism could incorporate both PN and NR results was based on the consideration of bound states (elliptic motions)  of compact binaries~\cite{Buonanno:1998gg,Buonanno:2000ef,Damour:2000we,Damour:2001tu,Damour:2012ky,Taracchini:2012ig}. 

Recently, a novel approach to the EOB description of binary systems was introduced based on 
the consideration of unbound, scattering states of binary black-hole systems~\cite{Damour:2016gwp}.
When considering non-spinning systems the key tool for this approach is the functional dependence of the center-of-mass frame scattering 
angle $\chi$ on the total energy $E$ and the total orbital angular momentum  $L$ of the binary system (also considered in the center-of-mass frame). If we consider binary systems of spinning bodies (taken, for simplicity, with parallel spins $S_1$ and $S_2$) the scattering angle $\chi$ will depend on four
variables:
\begin{eqnarray}
\label{chi_def_1}
\chi &=&\chi(E,L,S_1,S_2)\nonumber\\
&=& \chi_{\rm orb}(E,L) + \chi_{\rm spin}(E,L,S_1,S_2) \,.
\end{eqnarray}
In addition, Ref.~\cite{Damour:2014afa} showed how to use (in the case of non-spinning bodies) the (orbital) gauge-invariant scattering function 
$\chi_{\rm orb}(E,L)$, Eq. \eqref{chi_def_1}, to compare analytical results with numerical simulations of  hyperbolic encounters of binary black-hole systems.

 The (orbital) scattering function $\chi_{\rm orb}(E,L)$ was analytically computed at the second post-Minkowskian approximation in Refs. \cite{Westpfahl:1985} based on the 2PM equations of motion derived in Refs. \cite{Westpfahl:1979gu,Bel:1981}.
Within the PN approximation method  the scattering function is (when considering the {\it conservative} dynamics) formally expanded according to
\begin{eqnarray}
\chi(E, L, S_1, S_2) &=& \chi_{\rm orb}(E,L)\nonumber\\ 
&&+  \chi_{S_1}(E,L) S_1 + \chi_{S_2}(E,L) S_2 \nonumber\\ 
&&+ O( {\rm spin}^2) \,,
\end{eqnarray}
where, as indicated, we will limit our investigation to the linear-in-spin contributions, and where
\begin{eqnarray}
\label{chi_def_2}
\chi_{\rm orb}(E,L)&=& \chi^{\rm N}(E',L) + \frac{1}{c^2} \chi^{\rm 1PN}(E',L)\nonumber\\
&&+\frac{1}{c^4} \chi^{\rm 2PN}(E',L)+\frac{1}{c^6} \chi^{\rm 3PN}(E',L)\nonumber\\ 
&& +\frac{1}{c^8} \chi^{\rm 4PN}(E',L)+\ldots 
\end{eqnarray}
where $E'\sim E-Mc^2$ is a measure of the nonrelativistic energy content (see below for the definition of the exact energy variable 
$E'=\mu c^2 \bar E$ which we shall use) and where the leading-order, \lq\lq Newtonian," term,  $\chi^{\rm N}(E',L)$, is given by
\beq
\chi^{\rm N}(E',L)= 2 \, \arctan \left( \frac{G m_1 m_2}{  L}
\sqrt{\frac{\mu  }{2 E' }} \right) \,.
\eeq
Our notation here is
\beq \label{notation1}
M=m_1+m_2\,,\qquad \mu=\frac{m_1m_2}{(m_1+m_2)}\,,\qquad \nu=\frac{\mu}{M}\,,
\eeq
where $m_1$ and $m_2$ are the two masses of the components of the binary system.
So far  only the second post-Newtonian (2PN) approximation to $\chi(E,L)$ ($O(1/c^4)$),  Eq. \eqref{chi_def_2}, has been explicitly derived~\cite{Bini:2012ji}.
 
The aim of the present paper is: (i) to extend the analytical knowledge of the orbital scattering function   up to the 4PN $O(1/c^8)$ level;
and (ii) to compute the linear-in-spin contribution to $\chi$ to next-to-next-to-leading PN order. 
Our calculation will be entirely based on the EOB formalism. In particular, we shall make a crucial use of the recent derivation~\cite{Damour:2014jta,Jaranowski:2015lha} of the 4PN conservative dynamics in  Arnowitt-Deser-Misner (ADM) coordinates and on its transcription within the EOB formalism~\cite{Damour:2015isa}.
For earlier  partial 4PN results see Refs.~\cite{Foffa:2012rn,Jaranowski:2012eb,Jaranowski:2013lca,Bini:2013zaa}. See also~\cite{Bernard:2015njp,Damour:2016abl,Bernard:2016wrg} for a discussion of the harmonic coordinates counterpart of the 4PN dynamics.

Subtle issues in the computation of the scattering function arise at the 4PN level, because of the {\it nonlocal-in-time} character of the 
two-body action. In the Hamiltonian formalism the conservative two-body dynamics is described by an action of the type
\beq
\label{action_1}
S({\mathbf Q},{\mathbf P})=\int \left(P_i \dot Q^i - H_{\rm (tot)}[T; {\mathbf Q}(\cdot), {\mathbf P}(\cdot)] \right) dT\,,
\eeq
where ${\mathbf Q}$ and ${\mathbf P}$ are phase space variables describing the relative motion in the center-of-mass frame.
 It was found in Ref. \cite{Damour:2014jta} (and later confirmed in \cite{Bernard:2015njp}) that the total 4PN-level Hamiltonian  $H_{\rm (tot)}$ is the sum of a usual {\it local-in-time} Hamiltonian $H_{\rm (loc)}$ and of a nonlocal-in-time contribution $H_{\rm (nonloc)}^{\rm tail}$, i.e., 
\begin{eqnarray} \label{loc+nonloc}
H_{\rm (tot)}[T; {\mathbf Q}(\cdot), {\mathbf P}(\cdot)] &=& H_{\rm (loc)}(T; {\mathbf Q}(T), {\mathbf P}(T)) \nonumber\\ 
&+& H^{\rm tail}_{\rm (nonloc)}[T; {\mathbf Q}(\cdot), {\mathbf P}(\cdot)]\,. 
\end{eqnarray}
Here the notation $H_{\rm (loc)}(T; {\mathbf Q}(T), {\mathbf P}(T))$  represents a function of the phase space variables at time $T$, while the notation 
$H^{\rm tail}_{\rm (nonloc)}[T; {\mathbf Q}(\cdot), {\mathbf P}(\cdot)]$ represents, for a given time $T$, a functional of the entire time development of the phase space variables  $ {\mathbf Q}(T')$ and ${\mathbf P}(T')$ for $T'\not = T$.
More precisely the structure of $H_{\rm (nonloc)}^{\rm tail}$, which is directly related to tail-transported effects \cite{Blanchet:1987wq},
 reads (see Eqs. (3.4) and (3.6) of Ref. \cite{Damour:2015isa})
\begin{eqnarray}
\label{H_nonloc}
&& H_{\rm (nonloc)}^{\rm tail}[T; {\mathbf Q}(\cdot), {\mathbf P}(\cdot)]= -\frac{G^2M}{5c^8}I_{ij}^{(3)}(T)\,\times \nonumber\\
&&\qquad\qquad {\rm Pf}_{2s^{\rm phys}/c} \int \frac{dT'}{|T-T'|}I_{ij}^{(3)}(T')\,.
\end{eqnarray}
Here $I_{ij}^{(3)}(T)$ denotes the third derivative of the quadrupole moment of the system at time $T$ (expressed in terms  of the instantaneous values ${\mathbf Q}(T)$ and ${\mathbf P}(T)$) while $I_{ij}^{(3)}(T')$ denotes the corresponding quantity at a time $T'$ different from $T$ (expressed in terms of  ${\mathbf Q}(T')$ and ${\mathbf P}(T')$). In addition, the logarithmically divergent integral of $T'$ is defined by means of  a partie finie operation ${\rm Pf}$ using as a regularization time scale
the value $T_{\rm scale}=2s^{\rm phys}/c$, where $s^{\rm phys}$ is an arbitrary length scale. [The length scale $s^{\rm phys}$ enters $H_{\rm (loc)}$ as an infrared regularization scale and it enters $H_{\rm (nonloc)}^{\rm tail}$ as an ultraviolet regularization scale; the dependence on $s^{\rm phys}$ cancels between the two contributions.]  

In view of the dual structure of the action at the 4PN level, Eq. \eqref{action_1}, leading to a corresponding dual structure of the scattering function, 
\beq
\label{chi_def_3}
\chi(E,L)=\chi_{\rm loc}(E,L)+\chi_{\rm tail}(E,L)\,,
\eeq
our  computation of the scattering function will be done by combining two separate calculations. 

On the one hand, the PN-expansion of the first (local) contribution will be computed by standard Hamiltonian methods within the EOB transcription of the 4PN dynamics \cite{Damour:2015isa}
\begin{eqnarray}
\label{chi_PN}
\chi_{\rm loc}(E,L)&=&\chi^{\rm (N)}(\bar E,L) + \frac{1}{c^2} \chi^{\rm (1PN)}(\bar E,L)\nonumber\\
&+& \frac{1}{c^4} \chi^{\rm (2PN)}(\bar E,L)+\frac{1}{c^6} \chi^{\rm (3PN)}(\bar E,L)\nonumber\\ 
&+& \frac{1}{c^8} \chi^{\rm (4PN)}_{\rm loc}(\bar E,L)+O\left( \frac{1}{c^{10}}\right)\,.
\end{eqnarray}
On the other hand, the computation of the tail contribution to $\chi(E,L)$ [which is entirely at the 4PN level]
\beq
\chi_{\rm tail}(E,L)= \frac{1}{c^8} \chi^{\rm (4PN)}_{\rm tail}(\bar E,L) \,,
\eeq 
will be performed by two independent calculations.
One calculation will be based on a generalization of the \lq\lq time-localization"  method introduced in  Ref. \cite{Damour:2015isa} and further explained in \cite{Damour:2016abl}. Refs. \cite{Damour:2015isa,Damour:2016abl} showed, in the case of ellipticlike bound motions, that it was possible to replace the nonlocal-in-time Hamiltonian, Eq. \eqref{H_nonloc}, by a physically equivalent local-in-time Hamiltonian, $H_{\rm tail}^{\rm reduced}(T; {\mathbf Q}(T), {\mathbf P}(T))$, obtained in the form of an infinite expansion in eccentricity or equivalently radial momentum. Here, we shall show, for the first time, how to generalize such a time-localization method to the case of hyperboliclike, unbound motions.

As an independent confirmation of the value of $\chi_{\rm tail}(E,L)$ obtained from our hyperbolic-type localization method, we will also compute $\chi_{\rm tail}$ directly from
the nonlocal-in-time force  ${\pmb {\mathcal F}}_{\rm (nonloc)}^{\rm tail}(T)$ associated with the nonlocal Hamiltonian \eqref{H_nonloc} by using the corresponding evolution of the  Laplace-Lagrange-Runge-Lenz vector.
Let us mention in this respect that an analogous confirmation of the validity of the localization method of Ref. \cite{Damour:2015isa} has been  obtained, for the case of ellipticlike, bound motions, in several recent works. Indeed, the first {\it analytical}, 4PN-level, determination of the periastron advance of small-eccentrity motions
was obtained in  Ref. \cite{Damour:2015isa}, see Eqs. from (8.1b)  to (8.4) there, by using the  standard EOB-derived  expression \cite{Damour:2009sm} yielding periastron precession as function of the two EOB potentials $A(u;\nu)$ and  $\bar D(u; \nu)$. The value of the $\nu$-linear contribution (which crucially include the tail contributions) to the potential $\bar D(u; \nu)$ was first confirmed by the  analytical self-force  results of Refs. \cite{Bini:2015bfb,Hopper:2015icj} (which used the first law of mechanics for eccentric orbits, \cite{Tiec:2015cxa}) as well as by the high-accuracy dynamical numerical self-force calculations of Ref. \cite{vandeMeent:2016hel}. A further analytical confirmation of the periastron precession result of Ref. \cite{Damour:2015isa} was recently obtained by a direct dynamical computation  involving the nonlocal tail force ${\pmb {\mathcal F}}_{\rm (nonloc)}^{\rm tail}(T)$ in Ref. \cite{Bernard:2016wrg}. Finally, further confirmations of the reduced, time-localized action of \cite{Damour:2015isa} are contained in the recent work of Ref. \cite{Blanchet:2017rcn}.

\section{Notation and brief EOB reminders}

Before proceeding with the computation of the scattering angle, let us shortly recall here the EOB description of the orbital motion of a two-body system.
The total EOB Hamiltonian of the system is expressed as
\beq \label{Heob}
H({\mathbf Q}, {\mathbf P})=Mc^2 \sqrt{1+2\nu \left(\frac{H_{\rm eff}}{\mu c^2}-1 \right)}\equiv Mc^2 h\,,
\eeq
where we used the notation of Eq. \eqref{notation1}.

The ``effective Hamiltonian" $H_{\rm eff}$ entering Eq. \eqref{Heob} is given by
\begin{eqnarray}
\label{H_eff}
H_{\rm eff}&=& c^2 \sqrt{A\left( \mu^2 c^2 +{\mathbf P}^2 +\left(\frac{1}{B}-1\right)P_R^2+Q \right)}\nonumber\\
&+& \frac{G}{c^2 R^3} \left(g_S \mathbf{L} \cdot \mathbf{S} + g_{S_*} \mathbf{L} \cdot \mathbf{S_*} \right) \,.
\end{eqnarray}
Here ${\mathbf L}$ denotes the orbital angular momentum 
\beq
{\mathbf L}={\mathbf R}\times {\mathbf P}\,,
\eeq
(with magnitude $L= P_{\phi}$) while 
\beq
{\mathbf P}^2=P_R^2+\frac{{\mathbf L}^2}{R^2}\,. 
\eeq
The functions $A(R)$ and $B(R)$ entering $H_{\rm eff}$, Eq.\eqref{H_eff}, parametrize the effective metric
\beq
ds_{\rm eff}^2=-A(R)c^2 dT_{\rm eff}^2+B(R) dR^2 +R^2 d\phi^2\,,
\eeq
written in coordinates $(T_{\rm eff},R,\phi)$ and specialized here to equatorial motions. [$B$ is often replaced by another EOB function, $\bar D$,  defined so that $AB\bar D \equiv1$.] In addition, $Q$ in Eq. \eqref{H_eff} represents a post-geodesic (Finsler-type) contribution which is at least quartic in momenta. Finally, $\mathbf{S}$ and $\mathbf{S_*}$ denote the following combinations of the two spin vectors
\beq
\mathbf{S}= \mathbf{S_1} + \mathbf{S_2} \, ; \, \mathbf{S_*}= \frac{m_2}{m_1} \mathbf{S_1} + \frac{m_1}{m_2} \mathbf{S_2} \,,
\eeq
while $g_S$ and  $g_{S_*} $ are some corresponding gyro-gravitomagnetic ratios which will be defined below. In the following, we shall work to linear
order in the spins (so that we do not need to discuss the $O({\rm spin}^2)$ contributions to the EOB Hamiltonian), and we shall only
consider paralell (or anti-parallel) spins (i.e. nonprecessing binary systems).

It will be convenient to work with dimensionless rescaled independent variables, namely
\begin{eqnarray}
\label{resc_vars}
&& t=\frac{c^3T}{GM}\,,\quad r=\frac{c^2 R}{GM}=\frac{1}{u}\,,\nonumber\\
&& j=p_\phi=\frac{cP_\phi}{GM\mu}\,,\quad p_r=\frac{P_R}{\mu c}\,,
\end{eqnarray}
and dimensionless rescaled quantities, e.g.,
\beq
\hat H_{\rm eff}=\frac{H_{\rm eff}}{\mu c^2}\,, \quad  \hat Q=\frac{Q}{\mu^2 c^2}\,.
\eeq
Note that we use the notation $j$ (rather than the visually more delicate letter $\ell$) for the dimensionless {\it orbital} angular momentum. Hopefully, this
should not introduce confusion with the usual notation ${\mathbf J}={\mathbf L}+{\mathbf S}_1+{\mathbf S}_2$ for the total
conserved angular momentum  of the system. [Anyway, in most of the text we shall consider nonspinning bodies, and separately add, to linear
order, the effect of the spins.]
Moreover, we will often use units such that $G=1=c$, though we will sometimes put back the appropriate factors of $G$
or $c$, when it can illuminate the physical meaning of a final result.

Corresponding to the decomposition \eqref{loc+nonloc} of the Hamiltonian in local and nonlocal (or tail) parts, the EOB potentials $A$, $B$ and $Q$ 
admit similar decompositions, which we shall denote by the labels I (for the local piece) and II (for the nonlocal, tail, one), namely
\begin{eqnarray}
A(u;\nu)&=& A^{\rm I}(u;\nu) + A^{\rm II}(u;\nu) \,,\nonumber\\
\bar D(u;\nu)&=&\bar D^{\rm I}(u;\nu) + \bar D^{\rm II}(u;\nu)\,,\nonumber\\
\hat Q(u;\nu)&=&\hat Q^{\rm I}(u;\nu) + \hat Q^{\rm II}(u;\nu)\,,
\end{eqnarray}
Ref. \cite{Damour:2015isa}  has determined the values of the above potentials.
In particular, the values of the local contributions (which are valid for arbitrary motions, including hyperbolic ones) are
(see Eqs. (5.2), (5.3), (6.4) in \cite{Damour:2015isa})
\begin{eqnarray}
A^{\rm I}(u;\nu) &=&1-2u +\nu a_1(u)+\nu^2 a_2(u) \nonumber\\
\bar D^{\rm I}(u;\nu) &=& 1+\nu \bar d_1(u)+\nu^2 \bar d_2(u) \nonumber\\
\hat Q^{\rm I}(r,p_r;\nu)&=& [2 (4-3\nu)\nu u^2+(20\nu-83\nu^2+10\nu^3) u^3] p_r^4\nonumber\\
&+&\left(-\frac{9}{5}\nu-\frac{27}{5}\nu^2+6\nu^3\right) u^2 p_r^6\,,
\end{eqnarray}
with 
\begin{eqnarray}
a_1&=&   2 u^3+\left(\frac{94}{3}-\frac{41}{32}\pi^2\right) u^4+(a^{\rm I}_{5c} -a^{\rm I}_{5{\ln{}}}\ln(u))u^5\nonumber\\
a_2&=&  \left(\frac{41}{32}\pi^2-\frac{221}{6}\right) u^5\nonumber\\
\bar d_1&=&  6 u^2+52 u^3+(\bar d^{\rm I}_{4c} -\bar d^{\rm I}_{4{\ln{}}}\ln(u))u^4\nonumber\\
\bar d_2&=&   -6u^3+\left(\frac{123}{16}\pi^2-260\right) u^4\,,
\end{eqnarray}
and
\begin{eqnarray}
a^{\rm I}_{5c}&=& \left(\frac{2275}{512}\pi^2-\frac{4237}{60}-\frac{128}{5}\ln(s)\right)\nu\nonumber\\
&+&\left(\frac{41}{32}\pi^2-\frac{221}{6}\right)\nu^2\nonumber\\
a^{\rm I}_{5{\ln{}}}&=&\frac{128}{5}\nu \nonumber\\
\bar d^{\rm I}_{4c}&=& \left(\frac{7243}{45}-\frac{23761}{1536}\pi^2-\frac{1184}{15}\ln(s)\right)\nu\nonumber\\
&+& \left(\frac{123}{16}\pi^2-260\right)\nu^2\nonumber\\
\bar d^{\rm I}_{4{\ln{}}}&=&\frac{1184}{15}\nu \,.
\end{eqnarray}
Note that the scale $s$ entering above via its logarithm $\ln(s)$ is an adimensionalized version of the physical regularization length scale
$s^{\rm phys}$ mentioned in the Introduction. Namely, $s \equiv c^2 s^{\rm phys}/(GM)$.

By contrast, the nonlocal (or tail) contributions to the EOB potentials determined in Eqs. (7.12) of \cite{Damour:2015isa}, namely
\begin{eqnarray}
A^{\rm II}(u,\nu) &=& (a^{\rm II}_{5c} -a^{\rm II}_{5{\ln{}}}\ln(u))u^5 \nonumber\\
\bar D^{\rm II}(u,\nu) &=&  (\bar d^{\rm II}_{4c} -\bar d^{\rm II}_{4{\ln{}}}\ln(u))u^4 \nonumber\\
\hat Q^{\rm II}(u,p_r,\nu)&=& (q^{\rm II}_{43 c} -q^{\rm II}_{43{\ln{}}} \ln(u))u^3p_r^4\nonumber\\
&+&(q^{\rm II}_{62 c} -q^{\rm II}_{62{\ln{}}} \ln(u))u^2p_r^6 \nonumber\\
&+& O(u p_r^8)\,.
\end{eqnarray}
where
\begin{eqnarray}
a^{\rm II}_{5c}&=&\frac{128}{5}\left(\gamma +2\ln 2 +\ln s\right)\nu \nonumber\\
a^{\rm II}_{5{\ln{}}}&=& -\frac{192}{5}\nu\nonumber\\
\bar d^{\rm II}_{4c}&=& \left(-\frac{845}{5}+\frac{1184}{15}\gamma-\frac{6496}{15}\ln 2+\frac{2916}{5}\ln 3\right. \nonumber\\
&& \left. +\frac{1184}{15}\ln s\right) \nu\nonumber\\
\bar d^{\rm II}_{4{\ln{}}}&=&-\frac{592}{5}\nu \nonumber\\
q^{\rm II}_{43 c}&=& \left(-\frac{5608}{15}+\frac{496256}{45}\ln 2 -\frac{33048}{5}\ln 3  \right) \nu\nonumber\\
a^{\rm II}_{43{\ln{}}}&=&0 \nonumber\\
q^{\rm II}_{62 c}&=&\left(-\frac{4108}{15}-\frac{2358912}{25}\ln 2+\frac{1399437}{50}\ln 3\right. \nonumber\\
&&\left. +\frac{390625}{18}\ln 5 \right)\nu \nonumber\\
q^{\rm II}_{62{\ln{}}} &=& 0\,,
\end{eqnarray}
were derived in Ref. \cite{Damour:2015isa} by means of an expansion in powers of the eccentricity, and can thereby only
be used {\it in the vicinity of circular motions}. As we are here interested in motions that are very far from circular ones,
we shall not be able to use the latter results, and will have to treat the nonlocal contribution to $\chi$ by a different expansion
(essentially an expansion in {\it inverse} powers of the eccentricity).

\section{Contribution of the local conservative dynamics to the scattering function of nonspinning bodies} \label{sec:local}

As we work linearly in the spins (which only involve local-in-time interactions at the order we consider), and as it is enough to
work linearly in the 4PN-level nonlocal tail effects, we can decompose the full scattering function as a sum of separate contributions, namely
\begin{eqnarray}
\chi(E, L, S, S_*) &=& \chi_{\rm loc}(E,L) +  \chi_{\rm tail}(E,L) \nonumber\\
&+& \chi_{S}(E,L) S + \chi_{S_*}(E,L) S_* \nonumber\\
&+& O( {\rm spin}^2)\,.
\end{eqnarray}

In this section we compute the term $\chi_{\rm loc}(E,L)$ in Eq. \eqref{chi_def_3}, i.e., the contribution to the scattering function 
of {\it nonspinning bodies} coming from
the {\it local-in-time} part of the Hamiltonian. This calculation will be done by standard Hamiltonian methods, and will be 
conveniently performed within the EOB reformulation of the conservative dynamics, restricted  here to equatorial motions.

\subsection{EOB-derived integral expression for $\chi_{\rm loc}(E,L)$}

To compute the scattering angle we use the Hamilton-Jacobi approach. [We checked our results by using the alternative 
method introduced in Ref. \cite{Bini:2012ji}. We sketch this alternative approach in Appendix \ref{alternativemethod}.]
The EOB action takes the separated form
\begin{eqnarray}
S(T_{\rm eff},R,\phi; {\mathcal E}_{\rm eff}, P_\phi)&=&-{\mathcal E}_{\rm eff} T_{\rm eff} +P_\phi \phi \nonumber\\
&+& \int^R\,  dR\,  P_R(R,{\mathcal E}_{\rm eff},P_\phi) \,.
\end{eqnarray}
Here, $T_{\rm eff}$ is the coordinate time of the effective EOB metric, and
${\mathcal E}_{\rm eff}$ is the effective EOB energy, whose $\mu$-rescaled avatar 
$\hat {\mathcal E}_{\rm eff}\equiv  {\mathcal E}_{\rm eff}/(\mu c^2)$  is the energy variable which enters most naturally the EOB formalism. 
It is related to the total energy $E$ of the system via 
\begin{eqnarray}
E&=&Mc^2 \sqrt{1+2\nu \left(\hat {\mathcal E}_{\rm eff}-1  \right)}\nonumber\\
\hat {\mathcal E}_{\rm eff}&=& \frac{E^2-m_1^2c^4-m_2^2c^4}{2m_1 m_2 c^4}\,.
\end{eqnarray}
The  equation for the orbit is then obtained from
\beq
\frac{\partial S}{\partial P_\phi}=\phi_0={\rm constant}.
\eeq
As stated above, it is generally convenient to work with the dimensionless rescaled variables, Eq. \eqref{resc_vars}, notably:  
the dimensionless orbital angular momentum  $p_\phi=j$, the dimensionless rescaled radial momentum $p_r=P_R/\mu c$, and the dimensionless gravitational potential\footnote{In some cases, it will be useful to work with
dimensionful versions of the latter quantities, say $p_r^{\rm phys}=P_R/\mu$ (with dimension $[{\rm velocity}]$) and $u^{\rm phys}\equiv 1/r^{\rm phys}\equiv GM/R $ (with dimension $[{\rm velocity}]^2$).} $u\equiv 1/r\equiv GM/Rc^2 $. In terms of these, the orbital phase 
(from which we shall directly deduce the scattering angle $\chi$) is
given by an integral of the form
\begin{eqnarray}
\label{eq_phi_HJ}
\phi-\phi_0 &=&  -\int^u  \left( \frac{\partial  }{\partial j}\, p_r(u,\bar E,j) \right) \frac{du}{u^2}\nonumber\\
&\equiv &\int^u U(u,\bar E,j) du\,,
\end{eqnarray}
where
\beq
\label{U_def}
U(u,\bar E,j)\equiv -\frac{1}{u^2}  \frac{\partial  }{\partial j}\, p_r(u,\bar E,j)\,.
\eeq 
Here, we have introduced a new (dimensionless) energy variable $\bar E$
which is biunivocally related both to the
total relativistic energy of the system $E = M c^2 + \cdots$, with usual energy dimension, and to the 
the dimensionless relativistic effective energy $\hat {\mathcal E}_{\rm eff}=1+O(1/c^2)$. The definition of $\bar E$ is such that,
in the non-relativistic limit $c\to \infty$, it  coincides (modulo a factor $1/c^2$)\footnote{Only the product $ c^2 \bar E$ has a limit
when $c\to \infty$, but we find it more convenient to work with dimensionless quantities.}  with the fractional binding energy $(E- Mc^2)/(Mc^2)$. Namely, we define
\beq
\bar E\equiv \frac{1}2 \left(\hat {\mathcal E}_{\rm eff}^2-1 \right)\equiv \frac12 v_\infty^2\,.
\eeq
In the latter equation we have introduced, in addition to the notation $\bar E$, the related variable $v_\infty^2\equiv  2\bar E$.
[The notation $v_\infty^2$ has been chosen because, in the case of unbound motions, $\bar E>0$, the quantity $v_\infty\equiv \sqrt{2\bar E}$, indeed measures (in units of $c$) the relative velocity for infinite separations. In the case of unbound motions only $v_\infty^2$ will enter our results.]
Note that we have introduced so far, and will indifferently use below, several equivalent energy  variables: $E$, ${\mathcal E}_{\rm eff}$, $\bar E$ and $v_\infty$.

In order to get the explicit Hamilton-Jacobi integral form \eqref{eq_phi_HJ} of the orbital phase, we
need to express $p_r$ as a function of $u=1/r$, for given values of the energy and the angular momentum. This will follow by solving in $p_r$ the EOB energy conservation law, 
$E=H(r,p_r,j)$ or, equivalently,  $\hat {\mathcal E}_{\rm eff}^2=\hat H_{\rm eff}^2(r,p_r,j)$.  The explicit form of the latter conservation law reads
\begin{eqnarray}
\label{E_eff_pr}
\hat {\mathcal E}_{\rm eff}^2
&=& A(u;\nu)\left[1+j^2 u^2  + A(u;\nu)\bar D(u;\nu) p_r^2 \right. \nonumber\\
&&\left. +q_4(u;\nu) p_r^4 +q_6(u;\nu) p_r^6\right]\,.
\end{eqnarray}
We shall see below how one can,
 in a PN-expanded sense, iteratively solve Eq. \eqref{E_eff_pr} for $p_r^2$ to get an
 explicit form of the function   $p_r(u,\bar E,j)$.

Armed with the knowledge of the function $p_r(u,\bar E,j)$, and thereby of the function $U(u,\bar E,j)$, Eq. \eqref{U_def},
the scattering angle can then (in keeping with Eq. (5.65) in Ref. \cite{Bini:2012ji}) be expressed as the following definite integral
\beq
\label{scatt_angle}
\frac12 \left( \chi(\bar E,j)+\pi\right)=\int_0^{u_{\rm (max)}} U(u,\bar E,j) du\,,
\eeq
where $u_{\rm (max)}=u_{\rm (max)}(\bar E,j)=1/r_{\rm (min)}$ corresponds to the distance of closest approach between the two bodies. 
Note in passing that Eq. \eqref{scatt_angle} is closely similar to the Hamilton-Jacobi integral formula for the dimensionless periastron advance, namely  (as in Eq. (5.35) of Ref. \cite{Bini:2012ji})
\beq
\label{periastron_adv}
K\equiv 1+k =\frac{1}{\pi}\int_{u_{\rm (min)}}^{u_{\rm (max)}} U(u,\bar E,j) du\,,
\eeq
where $u_{\rm (min)}(\bar E,j)$ and $u_{\rm (max)}(\bar E,j)$  now respectively correspond to  apoastron and periastron   passages. 
The exact definitions of the functions $u_{\rm (min)}(\bar E,j)$ and $u_{\rm (max)}(\bar E,j)$ is that they are the two roots closest to zero of the EOB circular relation  
\beq
\label{EOB_circ}
\hat {\mathcal E}_{\rm eff}^2
=A(u;\nu)(1+j^2 u^2)\,.
\eeq

Note that if we define the function 
\beq
V(u; j, \bar E)=\int_0^{u} U(u,\bar E,j) du\,,
\eeq
we can write 
the following compact expressions 
\begin{eqnarray}
\frac12 \chi(\bar E,j) &=&  V(u_{\rm (max)}; \bar E,j)-\frac{\pi}{2}\,,\nonumber\\   
\pi \, K &=& V(u_{\rm (max)}; \bar E,j)-V(u_{\rm (min)}; \bar E,j)\,.
\end{eqnarray}
The first expression holds for $\bar E>0$, while the second applies in the case $\bar E<0$. Note that, when analytical continuing in $\bar E$ the definitions \eqref{def_u_min_max} 
of the functions 
 $u_{\rm (min)}(\bar E,j)$ and $u_{\rm (max)}(\bar E,j)$, one passes from a configuration where (when $\bar E<0$)  $0<u_{\rm (min)}<u_{\rm (max)}$ to a configuration where  (when $\bar E>0$)  $u_{\rm (min)}<0<u_{\rm (max)}$.
This shows that while $K$ has the nature of a {\it complete} hyperelliptic integral, $\chi$ has the nature of an {\it incomplete}  hyperelliptic integral.

\subsection{PN-expanding $\chi_{\rm loc}(E,L ; 1/c^2)$}

Let us now show how one can explicitly compute (as a PN-expansion) the function $U(u,\bar E,j)$, Eq. \eqref{U_def},
whose integral yields $\chi_{\rm loc}(E,L)$, Eq. \eqref{scatt_angle}. The first step is to compute the function $p_r(u,\bar E,j)$.
The latter function is obtained by iteratively solving (in successive powers of $1/c^2$) Eq. \eqref{E_eff_pr} for $p_r^2$. This yields 
\begin{eqnarray}
\label{p_r_expanded}
p_r^2&=&[p_r^2]^{(0)}+\eta^2[p_r^2]^{(2)}+\eta^4 [p_r^2]^{(4)}+ \eta^6[p_r^2]^{(6)}\nonumber\\
&+&\eta^8 [p_r^2]^{(8)}+O(\eta^{10})\,.
\end{eqnarray}
Here $\eta\sim 1/c$ is  a PN-order marker (to be taken to one at the end), and
\begin{eqnarray}
\label{vari_pr}
{}[p_r^2]^{(0)}&=& 2\bar E-j^2 u^2+2u\nonumber\\
{}[p_r^2]^{(2)}&=& 8\bar E u+8u^2-2j^2u^3 \nonumber\\
{}[p_r^2]^{(4)}&=& (24-12 \nu )\bar E u^2+(24 -14\nu )u^3+(6 \nu -4  )j^2 u^4\nonumber\\
{}[p_r^2]^{(6)}&=& (-32+24\nu)\nu \bar E^2 u^2+(64+60\nu^2-224\nu)\bar E u^3\nonumber\\
&+& \left[(32-24\nu)\nu j^2\bar E+64+\left( \frac{41}{32}\pi^2 -\frac{694}{3}\right)\nu\right. \nonumber\\
&+& \left. 36\nu^2\right] u^4+(-8+98\nu-30\nu^2) j^2 u^5\nonumber\\
&+& (-8+6\nu)\nu j^4 u^6\,,
\end{eqnarray}
with a similar expression for the 4PN term ${}[p_r^2]^{(8)}$, not displayed here for brevity.

A priori one also needs the PN-expansion\footnote{This PN-expansion is defined by keeping $\bar E$ fixed and expanding in powers of $1/c^2$.} of the functions  $u_{\rm (min)}(\bar E,j;1/c^2)$ and $u_{\rm (max)}(\bar E,j;1/c^2)$ which define the boundaries of the
integral expressions \eqref{scatt_angle} and \eqref{periastron_adv}. These expansions have the form
\begin{eqnarray}
\label{def_u_min_max}
u_{\rm min}\left(\bar E,j;\frac{1}{c^2}\right) &=&\frac{1-\sqrt{1+2j^2\bar E}}{j^2}+O\left(\frac{1}{c^2}  \right) \,,\nonumber\\
u_{\rm max}\left(\bar E,j;\frac{1}{c^2}\right) &=&\frac{1+\sqrt{1+2j^2\bar E}}{j^2}+O\left(\frac{1}{c^2}  \right) \,.  
\end{eqnarray}

The original expression \eqref{scatt_angle} for $\chi$ is given in terms of a convergent integral, having as upper limit the positive root closest to zero,
$u=u_{\rm max}(\bar E,j;1/c^2)$, of Eq. \eqref{EOB_circ}. When PN-expanding the integral expression \eqref{scatt_angle} for $\chi(\bar E,j;1/c^2)$ one should a priori PN-expand both the integrand  $U(u,\bar E,j ; 1/c^2)$, and the upper limit 
$u_{\rm max}(\bar E,j;1/c^2)$ (as per Eq. \eqref{def_u_min_max}). Such a formal PN expansion generates a sequence of {\it divergent} integrals on the interval $[0,u_{\rm max}^{\rm N}]$, where
\beq \label{umaxN}
u_{\rm max}^{\rm N}(\bar E,j)\equiv \frac{1+\sqrt{1+2j^2\bar E}}{j^2} \,, 
\eeq 
together with {\it formally infinite} contributions coming from the expansion of the upper limit. It was, however, shown in Ref. \cite{Damour:1988mr} that the correct value of the PN-expanded integral is recovered by: i) using as upper limit of the integral the Newtonian limit $u_{\rm max}^{\rm N}$, Eq. \eqref{umaxN}; ii) PN-expanding the integrand; iii) taking the Hadamard partie finie\footnote{For the considered integrals the original notion of Hadamard partie finie applies. Their partie finie are unambiguously defined, by contrast with
the partie finie entering the nonlocal action \eqref{H_nonloc} whose definition involves the choice of a regularization scale.} (Pf) of the so-generated divergent integrals. \\

\subsection{Explicit expression of the 4PN-level value of $\frac12\chi_{\rm loc}(\bar E,j; 1/c^2)$}

Most of the integrals generated by the PN-expansion technique explained in the previous subsection can be computed by standard techniques, except one of them whose integrand involves $\ln u$, which  will be separately discussed below.
To simplify the final expressions it is convenient to introduce the auxiliary variable $\alpha$, defined as 
\beq
\alpha \equiv  \frac{1}{\sqrt{2\bar E j^2}}=\frac{1}{v_\infty j}\,, 
\eeq
as well as the function $B(\alpha)$, defined as
\beq
B(\alpha) \equiv \arctan \alpha +\frac{\pi}{2}\,.
\eeq
Our PN counting is such that $ 2 \bar E= v_\infty^2=O(\frac1{c^2})$, while $j=O(c)$. In the Newtonian approximation ($c\to\infty$), the dimensionless variable $\alpha$ has a finite limit  that is linked to the
(Newtonian-level) eccentricity, say $e=\sqrt{1+ 2\bar E j^2}=\sqrt{1+\frac1{\alpha^2}}$. As, by contrast, $j$ grows proportionally to $c$,
the $n$-PN contribution to $\chi$ must scale as $\chi^{n \rm PN}=f_n(\alpha)/j^n$.

The result of the calculation of $\frac12 \chi_{\rm loc}$, PN-decomposed  as in Eq. \eqref{chi_PN}, is found to be the following  
\begin{eqnarray} \label{PNexpchi}
\frac12 \chi^{\rm (N)}(\bar E,j)&=&\arctan \alpha =B(\alpha)-\frac{\pi}{2}\nonumber\\
\frac12 \chi^{\rm (1PN)}(\bar E,j) &=&\frac{1}{j^2}\left[3B(\alpha) +  \frac{(3\alpha^2+2)}{ \alpha(1+\alpha^2)}  \right]   \nonumber\\
\frac12 \chi^{\rm (2PN)}(\bar E,j) &=& \frac{1}{j^4}\left[ C_{\rm (2PN)}^{B} B(\alpha)+C^{0}_{\rm (2PN)}   \right]\nonumber\\
\frac12 \chi^{\rm (3PN)}(\bar E,j)&=& \frac{1}{j^6}\left[ C_{\rm (3PN)}^{B} B(\alpha)+C^{0}_{\rm (3PN)}   \right]\nonumber\\
\frac12\chi_{\rm loc}^{\rm (4PN)}(\bar E,j)&=& \frac{1}{j^8}\left[ C_{\rm (4PN)}^{B} B(\alpha)+C^{0}_{\rm (4PN)}   \right]\nonumber\\
&&+I_\chi\,.
\end{eqnarray}
Here the $\alpha$-dependent coefficients entering $\frac12 \chi^{\rm (2PN)}$\footnote{Let us mention a typo in the corresponding 2PN term in
\cite{Bini:2012ji}: the sign of the coefficient of $\tilde E$ in the first Eq. (5.51)  (defining $A_{2a}$) 
 should be reversed, namely it should read $+ (5-2\nu)/2$.},
$\frac12 \chi^{\rm (3PN)}$ and $\frac12 \chi_{\rm loc}^{\rm (4PN)}$ are
\begin{widetext}
\begin{eqnarray}
C_{\rm (2PN)}^{B}&=&-\frac{3[2(5\alpha^2+1)\nu-5(1+7\alpha^2)] }{4\alpha^2}   \nonumber\\
C^{0}_{\rm (2PN)}&=& -\frac{[2(15\alpha^4+28\alpha^2+13)\nu-105\alpha^4-190\alpha^2-81] }{4 \alpha(1+\alpha^2)^2}\nonumber\\
C_{\rm (3PN)}^{B}&=& \left[\frac{1155}{4}+\left(\frac{615}{128}\pi^2-\frac{625}{2}\right)\nu+\frac{105}{8}\nu^2  \right]
+\frac{1}{\alpha^2}\left[ \frac{315}{4}+\frac{45}{4}\nu^2+\left(\frac{123}{128}\pi^2-109\right)\nu \right] 
+\frac{1}{\alpha^4}\left( -\frac32 \nu+\frac98 \nu^2 \right)\nonumber\\
C^{0}_{\rm (3PN)}&=& \frac{1}{(1+\alpha^2)^3}\left\{
\alpha^5 \left[   \frac{1155}{4}+\frac{105}{8}\nu^2+\left(\frac{615}{128}\pi^2-\frac{625}{2}\right)\nu  \right]
+\alpha^3 \left[ \frac{3395}{4}+\frac{185}{4}\nu^2+\left(-\frac{2827}{3}+\frac{1763}{128}\pi^2\right)\nu  \right]\right. \nonumber\\
&&  +\alpha \left[ \frac{3381}{4}+60\nu^2+\left(\frac{1681}{128}\pi^2-\frac{2939}{3}\right)\nu  \right] 
+\frac{1}{\alpha}\left[\frac{1221}{4}+\frac{135}{4}\nu^2+\left(-\frac{1153}{3}+\frac{533}{128}\pi^2\right)\nu\right]\nonumber\\
&&\left. +\frac{1}{\alpha^3}\left(\frac{64}{3}+\frac{55}{8}\nu^2-\frac{69}{2}\nu\right)
\right\}
\end{eqnarray}
and
\begin{eqnarray}
C_{\rm (4PN)}^{B}&=& \left( \frac{244}{\alpha^2}+\frac{1190}{3} +\frac{74}{5\alpha^4}    \right)\nu \ln s \nonumber\\
&& +
\left[\frac{225225}{64}-\frac{315}{16}\nu^3+\left(-\frac{7175}{256}\pi^2+\frac{132475}{96}\right)\nu^2
+\left(-\frac{1720271}{288}+\frac{2975735}{24576}\pi^2\right)\nu\right]\nonumber\\
&& +\frac{1}{\alpha^2}\left[ \frac{45045}{32}-\frac{525}{16}\nu^3+\left(\frac{35065}{32}-\frac{615}{32}\pi^2\right)\nu^2
+\left(\frac{257195}{4096}\pi^2-\frac{288805}{96}\right)\nu \right] \nonumber\\
&& +\frac{1}{\alpha^4}\left[ \frac{3465}{64}-\frac{225}{16}\nu^3
+\left(\frac{4827}{32}-\frac{369}{256}\pi^2\right)\nu^2+\left(\frac{33601}{8192}\pi^2-\frac{95681}{480}\right)\nu \right]\nonumber\\
&& +\frac{1}{\alpha^6}\left( \frac{9}{32}\nu-\frac{15}{16}\nu^3+\frac{27}{32}\nu^2 \right)\nonumber\\
C^{0}_{\rm (4PN)} &=& \left( \frac{5758}{15\alpha}+\frac{5606}{45\alpha^3}+\frac{64}{5(\alpha^2+1)}\alpha  \right)\nu \ln s \nonumber\\
&& +\frac{1}{\alpha}\left[ \frac{197785}{64}-\frac{315}{16}\nu^3+\left(\frac{43185}{32}-\frac{7011}{256}\pi^2\right)\nu^2
+\left(-\frac{7890919}{1440}+\frac{2731199}{24576}\pi^2\right)\nu \right] \nonumber\\
&& +\frac{1}{\alpha^3}\left[ \frac{37495}{64}-\frac{105}{4}\nu^3+\left(\frac{95791}{144}-\frac{8077}{768}\pi^2\right)\nu^2+\left(-\frac{3128449}{2160}+\frac{2292919}{73728}\pi^2\right)\nu \right]\nonumber\\
&& +\frac{1}{\alpha^5}\left( -\frac{2957}{160}\nu-\frac{113}{16}\nu^3+\frac{1061}{32}\nu^2\right)\nonumber\\
&& +\frac{2\alpha}{(\alpha^2+1)^4}-\frac{\alpha (4\nu-15)}{(\alpha^2+1)^3}\nonumber\\
&& +\frac{\alpha}{(1+\alpha^2)^2}\left[ \nu^2+\left(\frac{41}{32} \pi^2-\frac{187}{3} \right)\nu+\frac{155}{2}  \right]\nonumber\\
&& +\frac{\alpha}{(1+\alpha^2)}\left[  \left(\frac{365}{12}-\frac{41}{64} \pi^2\right)\nu^2
+\left(\frac{10189}{1024} \pi^2-\frac{59203}{120}   \right)\nu+\frac{1715}{4} \right]\,.
\end{eqnarray}
Finally, the last contribution $I_\chi$ to $\frac12 \chi_{\rm loc}^{\rm (4PN)}$ is defined as the following (Hadamard-regularized) integral
\begin{eqnarray}
\label{I_chi_def}
 I_\chi &\equiv& -\frac{16 j\nu   }{15}\, {\rm Pf}\, \int_0^{u_{\rm (max)}}  \frac{ u^4 \ln(u) (-74\bar E+37 j^2 u^2-62 u)}{(2\bar E-j^2 u^2+2 u)^{3/2}} du\,.
\end{eqnarray}
\end{widetext}

Using a suitable integration by parts, the singular integral $ I_\chi $, Eq. \eqref{I_chi_def},  can be recast as
\begin{eqnarray}
 I_\chi  &=& {\mathcal I}_\chi +\frac{(413\alpha^4-8\alpha^2-37)\nu}{15j^8\alpha^4} B(\alpha)\nonumber\\
&& +\frac{ \nu (4010\alpha^2+6195\alpha^4-1461)}{225\alpha^3 j^8 (\alpha^2+1)} \,,
\end{eqnarray}
where the first term is now defined in terms of the following {\it convergent} integral
\begin{widetext}
\beq
\label{matcal_I_chi_def}
{\mathcal I}_\chi \equiv\frac{16 \nu}{15(1+2\bar E j^2)}\, \int_0^{u_{\rm (max)}}  \frac{(u j^2-1)u^3(-310 u-296\bar E+222 j^2 u^2)\ln(u)}
{(2\bar E-j^2 u^2+2 u)^{1/2}}  du\,.
\eeq
\end{widetext}

\subsection{Computing the large-$j$ expansion of the logarithmic integral $ {\mathcal I}_\chi $}

The integral ${\mathcal I}_\chi $, Eq. \eqref{matcal_I_chi_def}, cannot be expressed in terms of elementary functions. Let us, however, explain how all the terms of the
 expansion of ${\mathcal I}_\chi(\bar E, j)$  in powers of $1/j$, at fixed energy, can be explicitly computed in terms of elementary functions.

First, let us replace the integration variable $u$ by a new variable $x$, and introduce a convenient expansion parameter $\epsilon$:
\beq
u=\frac{\sqrt{2\bar E}}{j}x\quad ; \quad \epsilon \equiv2\alpha=\frac{2}{v_\infty j} \,.
\eeq 
This yields the new form
\begin{widetext}
\beq
\label{s_I_chi}
{\mathcal I}_\chi=\frac{512 \nu}{15 \epsilon^4 j^8 (\epsilon^2+4)}  \, {\rm Pf}\, \int_0^{x_{\rm max}(\epsilon)} \frac{(2 x-\epsilon) x^3 (-155 x\epsilon-148+222 x^2)\ln \left(\frac{2x}{\epsilon j^2}\right)}{\sqrt{ 1-x^2+x\epsilon}}dx \,,
\eeq
\end{widetext}
where the upper limit is $x_{\rm max}(\epsilon)=\sqrt{1+\epsilon^2/4}+\epsilon/2=1+O(\epsilon)$. 
The large-$j$ expansion, at fixed $v_\infty$, then corresponds (modulo the overall $1/j^8$ factor and the nonlogarithmic
integral involving $\ln 1/j^2$) to a small-$\epsilon$ expansion. 
Using again the general result of Ref. \cite{Damour:1988mr} the expansion in powers of $\epsilon$ of ${\mathcal I}_\chi$ is simply obtained by: i) expanding the integrand of Eq. \eqref{s_I_chi} in powers of $\epsilon$ (which generates singular integrals); ii) using as upper limit
$x_{\rm max}(0)=1$; and iii) taking Hadamard's partie finie of the singular integrals.

Let us display, for illustration, the first three terms of the $\epsilon$-expansion of the ${\mathcal I}_\chi$ integrand
\begin{widetext}

\begin{eqnarray}
\label{integrando}
&& \frac{\nu}{j^8 }\left[\frac{256}{15\epsilon^4 }\frac{ x^4 (-148+222 x^2) \ln \left(\frac{2x}{\epsilon j^2}\right)}{\sqrt{1-x^2}}
+\frac{256}{15\epsilon^3 } \frac{x^3(155x^4-266 x^2+74)\ln \left(\frac{2x}{\epsilon j^2}\right)}{(1-x^2)^{3/2}} \right. \nonumber\\
&& \left. -\frac{64}{15\epsilon^2}\frac{x^4(680 x^2-703 x^4-310+222 x^6)\ln \left(\frac{2x}{\epsilon j^2}\right)}{(1-x^2)^{5/2}} 
+O\left(\frac{1}{\epsilon}\right)\right]\,.
\end{eqnarray}
The (finite part of the) integral between $0$ and $1$ of the latter, expanded, logarithm-dependent integrand can be computed by taking the $a\to 0$ limit of the  identity
\beq
\frac{d}{da }\int_0^1 x^a f(x) dx=\int_0^1  x^a \ln (x) f(x) dx\,.
\eeq
Actually, it is convenient to consider the following generalized integral
\begin{eqnarray}
\label{integrando2}
{\mathcal J}_\chi(a,b)&\equiv&\frac{\nu}{j^8 } \int_0^1 \left[\frac{256}{15\epsilon^4 }  x^4 (-148+222 x^2)  \left(\frac{2x}{\epsilon j^2}\right)^a (1-x^2)^{b-\frac12}
+\frac{256}{15\epsilon^3 }  x^3(155x^4-266 x^2+74) \left(\frac{2x}{\epsilon j^2}\right)^a (1-x^2)^{b-\frac32 }  \right. \nonumber\\
&& \left. -\frac{64}{15\epsilon^2} x^4(680 x^2-703 x^4-310+222 x^6) \left(\frac{2x}{\epsilon j^2}\right)^a (1-x^2)^{b-5/2} +O\left(\frac{1}{\epsilon}\right)\right] dx\,,
\end{eqnarray}
\end{widetext}
in which the powers of the singular denominators have also been shifted by $-b$.
All the integrals entering ${\mathcal J}_\chi(a,b)$, Eq. \eqref{integrando2}, can be trivially computed in terms  of Euler's beta function. Finally differentiating the result with respect to $a$ and taking the limits $a\to 0$ and $b\to 0$ we get (restoring $\bar E$ in lieu  of $\epsilon$)
\begin{eqnarray} \label{jexpchi}
{\mathcal I}_\chi(\bar E,j)&=&\frac{\nu \bar E^2}{j^4}{\mathcal I}_4+
\frac{\nu \bar E^{3/2}}{j^5}{\mathcal I}_5+\frac{\nu \bar E }{j^6}{\mathcal I}_6\nonumber\\
&+&
\frac{\nu \bar E^{1/2}}{j^7}{\mathcal I}_7+
\frac{\nu }{j^8}{\mathcal I}_8
 +O\left(\frac{\bar E^{(-1/2)}}{j^9}\right)\,,
\end{eqnarray}
where
\begin{eqnarray}
{\mathcal I}_4&=& -\frac{37\pi }{5}\left[-3+\ln\left(\frac{4j^4}{\bar E^2}\right)\right]\nonumber\\
{\mathcal I}_5&=& \frac{448   \sqrt{2}}{675} \left[-218+840\ln(2)-105\ln\left(\frac{4j^4}{\bar E^2}\right)\right]\nonumber\\
{\mathcal I}_6&=& -\frac{\pi}{3} \left[-758+183\ln\left(\frac{4j^4}{\bar E^2}\right)\right] \nonumber\\
{\mathcal I}_7&=&  \frac{32  \sqrt{2} }{225 }\left[+8760\ln(2)-538-1095\ln\left(\frac{4j^4}{\bar E^2}\right)\right]\nonumber\\
{\mathcal I}_8&=&-\frac{\pi }{180}\left[-57362+8925\ln\left(\frac{4j^4}{\bar E^2}\right)\right] \,.
\end{eqnarray}

Here, for simplicity, only the first terms of the expansion have been displayed, but our method allows one to compute
as many terms as one wishes.

The large-$j$ expansion \eqref{jexpchi} (at fixed energy) is an expansion in powers of $\alpha$, and is equivalent to an
expansion in inverse powers of the (Newtonian-level) eccentricity $e=\sqrt{1+ 2\bar E j^2}=\sqrt{1+\frac1{\alpha^2}}$.
In addition, one can analytically compute the value of ${\mathcal I}_\chi(\bar E,j)$ at the other boundary (besides $\alpha=0$,
corresponding to $e=\infty$)
of the family of unbound, hyperbolic motions, namely $\alpha=\infty$, corresponding to the marginally bound case $\bar E=0$, i.e. the parabolic limit $e=1$.
One finds
\beq
{\mathcal I}_\chi(\bar E=0,j)=\frac{\nu}{j^8} \frac{\pi}{90} (47867 - 35700 \ln (2j^2))\,.
\eeq

\subsection{Final result for the large-$j$ expansion of the  local 4PN contribution to $\chi$}

We have given in Eqs. \eqref{PNexpchi} above the explicit expression of the 4PN-accurate value of the local contribution to
the scattering angle. Our result was fully explicit, except for the last, integral, contribution $I_\chi$ to the 4PN-level contribution 
$\frac12\chi_{\rm loc}^{\rm (4PN)}(\bar E,j)$ in Eq. \eqref{PNexpchi}. We have then transformed $I_\chi$, Eq. \eqref{I_chi_def},
into the simpler integral ${\mathcal I}_\chi$, Eq. \eqref{matcal_I_chi_def}. 
In the previous subsection, we have computed the large-$j$ expansion of ${\mathcal I}_\chi$.
Let us insert the latter result in our previous results to give the beginning of the  large-$j$ expansion (at fixed energy)
of  $\frac12\chi_{\rm loc}^{\rm (4PN)}(\bar E,j)$.

Both ${\mathcal I}_\chi$ and $I_\chi$ start at order $1/j^4$ in their large-$j$ expansion .
Let us first display the large-$j$ expansion of the (exactly known) part of $\chi_{\rm loc}^{\rm 4PN}/2$, namely the part that does not
include ${\mathcal I}_\chi(\bar E,j)$. It reads
\begin{eqnarray} \label{chi4PNminuscalI}
\frac{\chi^{\rm 4PN}_{\rm loc}}{2}-{\mathcal I}_\chi&=& \frac{v_\infty^6}{j^2}\left[  \frac{27}{64}\pi\nu^2+\frac{9}{64}\pi\nu-\frac{15}{32}\pi\nu^3  \right]\nonumber\\
&+&\frac{v_\infty^5}{j^3}\left[  -8\nu^3-\frac{91}{5}\nu+34\nu^2 \right]\nonumber\\
&+& \frac{v_\infty^4}{j^4}\left[-\frac{225}{32}\pi\nu^3+\left(\frac{4827}{64}\pi-\frac{369}{512}\pi^3\right)\nu^2 \right. \nonumber\\
&+&\left(-\frac{19373}{192}\pi+\frac{33601}{16384}\pi^3+\frac{37}{5}\pi\ln(s)\right)\nu\nonumber\\
&+&\left.  \frac{3465}{128}\pi   \right]+O\left(\frac{1}{j^5}\right)\,.
\end{eqnarray}
On the other hand, we have computed above the large-$j$ expansion of ${\mathcal I}_{\chi}$. Its first term reads
\begin{eqnarray} \label{calI}
{\mathcal I}_{\chi}&=& -\frac{37\pi\nu v_\infty^4}{20 j^4}\left[\ln\left(\frac{4j^4}{\bar E^2}\right)-3\right]
+O\left(\frac{1}{j^5}\right)\,.
\end{eqnarray}
Combining \eqref{chi4PNminuscalI} and \eqref{calI} finally yields the following large-$j$ expansion of the local 4PN scattering angle
\begin{eqnarray}
\label{chi_loc_j_gg1}
\frac{\chi^{\rm 4PN}_{\rm loc}}{2}&=& \frac{v_\infty^6}{j^2}\left[  \frac{27}{64}\pi\nu^2+\frac{9}{64}\pi\nu-\frac{15}{32}\pi\nu^3  \right]\nonumber\\
&+&\frac{v_\infty^5}{j^3}\left[  -8\nu^3-\frac{91}{5}\nu+34\nu^2 \right]\nonumber\\
&+& \frac{v_\infty^4}{j^4}\left[-\frac{225}{32}\pi\nu^3+\left(\frac{4827}{64}\pi-\frac{369}{512}\pi^3\right)\nu^2 \right. \nonumber\\
&+& \left(-\frac{91537}{960}\pi +\frac{33601}{16384}\pi^3 +\frac{37}{5}\pi\ln\left(\frac{v_\infty s}{2j}\right)\right)\nu\nonumber\\
&+&\left. \frac{3465}{128}\pi   \right]+O\left(\frac{1}{j^5}\right)\,.
\end{eqnarray}

\subsection{Schwarzschild limit}

As a check on the above comparable-mass ($\nu = O(1)$) results we have also computed the PN expansion of the extreme-mass-ratio
($\nu \to 0$) scattering angle. This is obtained 
by considering the scattering angle of a test particle of mass $\mu \to 0$ around a Schwarzschild black hole of mass $M$. It is given in terms of incomplete elliptic integrals
\cite{Chandrasekhar:1985kt} (Eq. (203), page 215 there) and reads
\beq
\frac{\chi_{s}}{2}= \frac{\kappa}{\sqrt{e_p u_p}}\left[K(\kappa)-F\left( \sqrt{\frac{e_p-1}{2e_p}},\kappa \right)  \right]-\frac{\pi}{2}\,.
\eeq
where
\beq
\kappa = 2 \sqrt{\frac{e_pu_p}{1-6u_p+2e_p u_p}}\,.
\eeq
Here $u_p$ and $e_p$ are  used to parametrize the conserved 
specific test-particle energy $E_s ={\mathcal E}^{\rm particle}/\mu$ and angular momentum $j_s= P_\phi^{\rm particle}/\mu$, according to the defining equations
\begin{eqnarray}
E_s &=& \sqrt{\frac{(1-2u_p)^2-4e_p^2u_p^2}{1-3u_p-e_p^2 u_p}}\nonumber\\
j_s &=& \frac{1}{\sqrt{u_p(1-3u_p-e_p^2 u_p)}}\,.
\end{eqnarray}
As the $\nu\to 0$ limits of the  EOB quantities $\hat {\mathcal E}_{\rm eff}$ and $j$ are equal (by construction of the EOB formalism) to $E_s$ and $j_s$, it is enough to re-express the Schwarzschild scattering angle $\chi_{s}(E_s,j_s))$ in terms of $j_s$ and of the Schwarzschild analogue of the two-body variable $\alpha$ used above, namely
\beq
\alpha_{s} \equiv\frac{1}{j_s\sqrt{E_s^2-1}} = \lim_{\nu \to 0} \alpha(\bar E, j)\,.
\eeq
This leads, suppressing for clarity the $s$ subscript on the independent variables and displaying only the first few terms, to
\begin{eqnarray}
\frac{\chi^{s}(j,\alpha)}{2}&=& {\rm arctan}(\alpha) \nonumber\\
&+& \left[\frac32 \pi+3{\rm arctan}(\alpha)+\frac{(3\alpha^2+2)}{\alpha (\alpha^2+1)} \right]\frac{\eta^2}{j^2}\nonumber\\
&+& \left[ \left(\frac{15}{4\alpha^2} +\frac{105}{4}\right){\rm arctan}(\alpha)+\frac{15}{8 \alpha^2}\pi\right.\nonumber\\
&+& \frac{81}{4 \alpha(\alpha^2+1)^2}+\frac{95\alpha}{2(\alpha^2+1)^2} +\frac{105}{8}\pi \nonumber\\
&+&\left. \frac{105\alpha^3}{4(\alpha^2+1)^2 } \right] \frac{\eta^4}{j^4}+ \ldots \,.
\end{eqnarray}
Here we have indicated the PN order by the formal PN-expansion parameter $\eta \sim 1/c$. [As explained above, the rule for doing so is to count $\alpha$ as being independent of $\eta$, while
$1/j$ is $O(\eta)$.]

Let us also exhibit the beginning of the expansion  of $\chi_s (E_s,j_s)$ in the limit where $\alpha_s \to 0$, keeping fixed the value of $E_s$.
[This is also the limit $1/j_s \to 0$ with fixed $E_s$.]
The result can be displayed in a more compact manner by re-expressing it in terms of both $1/j_s$ and $\alpha=\alpha_s= 1/(j_s\sqrt{E_s^2-1})$:
\begin{eqnarray}
\frac12  \chi_{s}(E_s,j_s) &=&
\frac12  \chi_{s}^{(0)}+\frac12  \chi_{s}^{(2)}\frac{\eta^2}{j^2} +\frac12  \chi_{s}^{(4)}\frac{\eta^4}{j^4} \nonumber\\
&+&\frac12  \chi_{s}^{(6)}\frac{\eta^6}{j^6} + \frac12  \chi_{s}^{(8)}\frac{\eta^8}{j^8}\nonumber\\
& +& O\left(\frac{\eta^{10}}{j^{10}}\right)\,,
\end{eqnarray}
where 
\begin{eqnarray}
\frac12  \chi_{s}^{(0)}&=&\alpha-\frac13 \alpha^3+\frac15 \alpha^5-\frac17 \alpha^7+\frac19\alpha^9 +O(\alpha^{11}) \nonumber\\
\frac12  \chi_{s}^{(2)}&=& \frac{2}{\alpha}+\frac32 \pi +4\alpha-2\alpha^3+\frac{8}{5}\alpha^5-\frac{10}{7}\alpha^7 +O(\alpha^{9}) \nonumber\\
\frac12  \chi_{s}^{(4)}&=&  \frac{15}{8 \alpha^2}\pi+\frac{24}{\alpha}+\frac{105}{8}\pi+32\alpha-16\alpha^3\nonumber\\
&&+\frac{96}{7}\alpha^5  +O(\alpha^{7}) \nonumber\\
\frac12  \chi_{s}^{(6)}&=&  \frac{64}{3\alpha^3} + \frac{315}{8 \alpha^2}\pi+\frac{320}{\alpha}\nonumber\\
&&+\frac{1155}{8}\pi+320\alpha-\frac{448}{3}\alpha^3 +O(\alpha^{5})  \nonumber\\
\frac12  \chi_{s}^{(8)}&=&   \frac{3465}{128\alpha^4}\pi+\frac{640}{\alpha^3}+\frac{45045}{64 \alpha^2}\pi+\frac{4480}{\alpha}+\frac{225225}{128}\pi \nonumber\\
&+&  3584\alpha +O(\alpha^{2}) \,.
\end{eqnarray}
Note that, in this expansion, a combination of the type $\alpha^p/j^q$ is of order $1/j^{p+q}$ in the large-$j$ expansion that we are considering.

We have checked that the $\nu \to 0$ limit of our comparable-mass results for $\chi$ given in the previous subsections agree
with the Schwarzschild limit.

\section{On the two approaches to the tail contribution to the conservative dynamics} \label{sectail1}

As recalled in the Introduction, it was shown in Ref. \cite{Damour:2014jta} (see Eq. (5.6) there) that the 4PN Hamiltonian is the sum of a local contribution, $H_{\rm (local)}$, and of a tail  contribution originally given as the following  nonlocal-in-time Hamiltonian, $H^{\rm tail}_{\rm (nonloc)}$
\begin{eqnarray}
\label{H_tail_I3_I3}
&& H_{\rm (nonloc)}^{\rm  tail }(T,{\mathbf Q},{\mathbf P};s)=\nonumber\\
&&-\frac{G^2M }5 \eta^8 \dddot I_{ij}(T)\,\,  {\rm Pf}_{2s/c}\int_{-\infty}^\infty \frac{d\tau }{|\tau|}\dddot I_{ij}(T+\tau)\,, 
\end{eqnarray}
where, we recall, $I_{ij}$ denotes  the Newtonian
quadrupole moment of the binary system in the center-of-mass frame, namely
\beq
\label{quadrup}
I_{ij}= \mu \left(Q^i Q^j -\frac13 Q^2 \delta^{ij}  \right)\equiv   \mu Q^{\langle i} Q^{j\rangle} \,.
\eeq
Here  $X^{\rm STF}\equiv X^{\langle ij\rangle}$ denotes the symmetric-trace-free part of the 2-tensor $X^{ij}$; similarly, $X^{\rm TF}$ denotes its trace-free part.

There are two ways to deal with the tail contribution to the 4PN  dynamics. On the one hand, the traditional one would be to compute the nonlocal tail force
induced by $H_{\rm (nonloc)}^{\rm  tail }$, and  to estimate how it modifies the dynamical effects linked to the local part of the 4PN-accurate dynamics.
On the other hand, a new approach has been introduced in Refs. \cite{Damour:2015isa,Damour:2016abl} in the case of ellipticlike motions. This second approach consists  in \lq\lq localizing," i.e., order-reducing, the nonlocal tail Hamiltonian $H_{\rm (nonloc)}^{\rm  tail }$, Eq. \eqref{H_tail_I3_I3}, to a physically equivalent {\it local} tail Hamiltonian $H_{\rm (loc)}^{\rm  tail }$.

The basic quantity entering the traditional approach is the nonlocal tail force ${\pmb {\mathcal F}}_{\rm (nonloc)}$.
Its value  is obtained by varying the following nonlocal action 
\begin{eqnarray}
\label{action_1bis}
&& S({\mathbf Q},{\mathbf P})=\\
&& \int \left(P_i \dot Q^i - H_{\rm loc}({\mathbf Q} , {\mathbf P} ) -H_{\rm (nonloc)}^{\rm  tail } [{\mathbf Q}(\cdot ) , {\mathbf P}(\cdot ) ] \right) dT\,.\nonumber
\end{eqnarray}
After some integration by parts, the equations of motion following from this action read (as indicated on pages 5 and 10 of Ref.  \cite{Damour:2014jta})
\begin{eqnarray}
\dot Q^i&=& \frac{\partial H_{\rm (loc)}}{\partial P_i}\nonumber\\
\dot P_i&=& -\frac{\partial H_{\rm (loc)}}{\partial Q^i}-\frac{4 G^2M }5    \mu Q^j(t)  \,\,  {\rm Pf}_{2s/c} \int d\tau  \frac{ I^{(6)}_{ij}(T+\tau) }{|\tau|}\nonumber\\
&\equiv & -\frac{\partial H_{\rm (loc)}}{\partial Q^i}+ {\mathcal F}^i_{\rm (nonloc)}(T)\,,
\end{eqnarray}
where 
\begin{eqnarray}
\label{F_non_loc}
&& {\mathcal F}^i_{\rm (nonloc)}(T)=-\frac{4G^2M }5 \eta^8    \mu Q^j(T)\times \nonumber\\
&&  \qquad {\rm Pf}_{2s/c} \int_{-\infty}^\infty d\tau  \frac{ I^{(6)}_{ij}(T+\tau) }{|\tau|}\,.
\end{eqnarray}
Note that, as pointed out in  \cite{Damour:2014jta}, the sum of the acausal time-symmetric nonlocal tail force, Eq. \eqref{F_non_loc}, and of the acausal time-antisymmetric nonlocal radiation reaction force (written on page 10 of  \cite{Damour:2014jta}), namely
\begin{eqnarray}
&& {\mathcal F}^i_{\rm (RR)}(T)=-\frac{4 G^2M }5   \mu q^j(T)  \times \nonumber\\
&&  \,\,  {\rm Pf}_{2s/c} \int_{0}^\infty \frac{ d\tau} {|\tau|} [I^{(6)}_{ij}(T-\tau)-I^{(6)}_{ij}(T+\tau)]\,,
\end{eqnarray}
yields the causal tail-transported nonlocal force first derived in Ref. \cite{Damour:1988mr} 
\begin{eqnarray}
&& {\mathcal F}^{i\,\rm tail,t-even}_{\rm (nonloc)} (T)+ {\mathcal F}^{i\,\rm t-odd}_{\rm (RR)}(T)= \nonumber\\
&&  -\frac{8 G^2M }5   \mu Q^j(T) \,   {\rm Pf}_{2s/c} \int_0^{+\infty} d\tau  \frac{ I^{(6)}_{ij}(T-\tau) }{|\tau|}\,.
\end{eqnarray}

Let us now generalize the time-localization procedure of Refs. \cite{Damour:2015isa,Damour:2016abl} from the ellipticlike case to the present hyperboliclike one.

Following  Refs. \cite{Damour:2015isa,Damour:2016abl}, $H_{\rm (loc)}^{\rm  tail }$  can be simply obtained by replacing the occurrences of 
the time-displayed phase-space arguments ${\bf Q}(T+\tau)$, ${\bf P}(T+\tau)$ in Eq.  \eqref{H_tail_I3_I3} by the corresponding solutions of the (Newtonian level) Hamilton equations of motion. [As shown in \cite{Damour:2015isa,Damour:2016abl} this a priori forbidden use of the equations of motion in an action is valid modulo a nonlocal-in-time transformation of the phase space variables.]

Let us introduce a convenient notation to define this time-localization procedure. 
Given some Hamiltonian flow in phase space (that we can take here to be simply the Newtonian level flow), and given the phase space position, ${\mathbf Q}={\mathbf Q}(T)$ and ${\mathbf P}={\mathbf P}(T)$ at a given time $T$, we denote  by  $\llbracket {\mathbf  Q} \rrbracket_{_{({\mathbf Q},{\mathbf P})}}^{ (T+\tau)}$, $\llbracket {\mathbf  P} \rrbracket_{_{({\mathbf Q},{\mathbf P})}}^{ (T+\tau)}$ the solution  of Hamilton's evolution equations at time $T+\tau$, which takes as initial values at time $T$ the given data  ${\mathbf Q}(T)$ and ${\mathbf P}(T)$. For any phase space function $F({\mathbf Q},{\mathbf P})$ we then correspondingly denote the value of $F$ at $\llbracket {\mathbf  Q} \rrbracket_{_{ {\mathbf Q},{\mathbf P} }}^{ (T+\tau)}$, $\llbracket {\mathbf  P} \rrbracket_{_{ {\mathbf Q},{\mathbf P} }}^{ (T+\tau)}$
simply as $\llbracket F\rrbracket_{_{{\mathbf Q},{\mathbf P}}}^{ (T+\tau)}$.
With this notation the time-localized  version of the tail action reads
\begin{eqnarray}
\label{H_tail_loc_time}
&& H_{\rm (loc)}^{\rm  tail }(T,{\mathbf Q},{\mathbf P};s)=-\frac{G^2M }{5c^8}   I_{ij}^{(3)}({\mathbf Q}(T),{\mathbf P}(T))\,\,\times \nonumber\\
&&\qquad   {\rm Pf}_{2s/c}\int_{-\infty}^\infty \frac{d\tau }{|\tau|}
\llbracket    I_{ij}^{(3)} \rrbracket_{_{{\mathbf Q}(T),{\mathbf P}(T)}}^{ (T+\tau)}\,.
\end{eqnarray}
Here, the notation $I_{ij}^{(3)}({\mathbf Q},{\mathbf P})$ denotes the order-reduced third time derivative of the quadrupole moment, i.e., the function of ${\mathbf Q}$ and ${\mathbf P}$  obtained by using Hamilton's equations of motion to reduce higher derivatives in $\dddot I_{ij}$. 

Refs. \cite{Damour:2015isa,Damour:2016abl}  showed how to explicitly compute the time-localized Hamiltonian, Eq. \eqref{H_tail_loc_time}, in the case of  ellipticlike, bound motions. This was done by using both Delaunay variables and the Delaunay averaging technique (to eliminate time-periodic terms). The resulting action was obtained as a power series in the eccentricity $e$, in a neighborhood of circular motion.

In the present paper we shall show how to generalize the time-localization technique of  Refs. \cite{Damour:2015isa,Damour:2016abl}
to hyperboliclike motions, and to deduce from it the tail contribution to the scattering function. Our result will be expressed 
as  a power series in the reciprocal of the eccentricity $1/e$, in a neighborhood of $e=\infty$, which corresponds to straight line motion.
[In addition, we shall show how to compute the value of $H_{\rm (loc)}^{\rm  tail }$ over the full tange of hyperbolic motions.]
We have explicitly checked that, at leading order in $1/e$, the so-deduced result agrees with a direct computation based on
the secular evolution of the Newtonian-conserved Laplace-Lagrange-Runge-Lenz, eccentricity vector under the action of the 
nonlocal, tail  4PN-level force ${\pmb{ \mathcal F}}^i_{\rm (nonloc)}$, Eq. \eqref{F_non_loc}. To relieve the tedium, we present this
alternative calculation in Appendix \ref{RungeLenz}.

\section{Link between the tail contribution $\chi^{\rm  tail }$ to scattering and the time-integrated tail Hamiltonian $\int dt H^{\rm  tail }$ } \label{sectail2}

In this section we show how to explicitly compute the tail contribution $\chi^{\rm  tail }(\bar E, j)$ to the scattering function from the knowledge of the time-integral (from $- \infty$ to $+\infty$) of the tail Hamiltonian $H^{\rm  tail }$. We shall assume that we work with the
time-localized version, $H^{\rm  tail }_{\rm (loc)}$, of the nonlocal tail Hamiltonian, as defined in the previous section, 
Eq. \eqref{H_tail_loc_time}. [In the next section we shall show
how to explicitly compute the time-localized $H^{\rm  tail }_{\rm (loc)}$.]

As $H_{\rm (loc)}^{\rm  tail }$ is of order $O(1/c^8)$ one can compute its contribution to $\chi$ simply by considering the Hamiltonian $H^{\rm N}+H_{\rm (loc)}^{\rm  tail }$, where $H^{\rm N}$ is the Newtonian-order Hamiltonian, all the other PN contributions  having been already incorporated in the local computation of the previous section.

We then start from an ordinary Hamiltonian  of the form
\beq
H({\mathbf q},{\mathbf p})=H_{\rm N}({\mathbf q},{\mathbf p})+H^{\rm  tail }({\mathbf q},{\mathbf p})\,.
\eeq
It will be convenient to work with rescaled phase-space variables, notably $r=R/GM$, $p_r=P_R/\mu$, and $p_\phi =P_\phi/\mu$,
and the rescaled time $t=T/GM$, while leaving the Hamiltonian unrescaled. [We use here $c=1$.]
Actually, in order to simplify things, we shall consider in this section that we use units such that $M=1$ (in addition to using
$G=c=1$).  In such units, one only needs to keep track of the power of $\mu =\nu $ entering various quantities.

In terms of such units, the lowest-order (here taken as Newtonian-order) Hamiltonian reads
\beq
H_{\rm N}=\nu \left[ \frac12 \left(p_r^2+\frac{p_\phi^2}{r^2}\right)-\frac{1}{r} \right]\,.
\eeq
The second, tail contribution, is here assumed to be  an ordinary (time-local) Hamiltonian, simply denoted by $H^{\rm  tail }({\mathbf q},{\mathbf p})$\footnote{The formula we shall arrive at for the correction to the scattering function induced by an additional contribution  to the Hamiltonian is actually very general and applies to the case $H({\mathbf q},{\mathbf p})=H^0({\mathbf q},{\mathbf p})+\epsilon H^1({\mathbf q},{\mathbf p}).
$ }. In the following we will denote $p_\phi=j$.

Let us recall the general Hamilton-Jacobi derived formula 
\beq
\label{eq_for_chi}
\chi(\bar E,j) =- \frac{\partial}{\partial j} \int p_r(\bar E,j,r) dr
\eeq
where the function $p_r(\bar  E,j,r)$ is defined by 
solving the energy conservation law
\beq
\bar E=\frac{H(r,p_r,j)}{\nu} = \frac12 \left(p_r^2+\frac{p_\phi^2}{r^2}\right)-\frac{1}{r} + \frac{H^{\rm tail}(r,p_r,j)}{\nu}. 
\eeq
When working to linear order in $H^{\rm  tail }$  the solution of the equation
\beq
p_r^2 =   2\bar E-\frac{j^2}{r^2}+\frac{2}{r}-2 \frac{H^{\rm tail}(r,p_r,j)}{\nu}
\eeq
is
\beq
\label{pr_eq}
p_r = p_r^{(0)} -\frac{1}{p_r^{(0)} } \frac{H^{\rm  tail } (r,p_r^{(0)},j)}{\nu}\,,
\eeq
where
\beq
p_r^{(0)}=\pm \sqrt{2\bar E-\frac{j^2}{r^2}+\frac{2}{r}}\,.
\eeq
Inserting Eq. \eqref{pr_eq} in Eq. \eqref{eq_for_chi} we find
\beq
\chi(\bar E, j) =
\chi_{\rm N}(\bar E, j)+\chi^{\rm tail} (\bar E, j)
\eeq
where
\beq
\chi_{\rm N}(\bar E, j)={\rm arctan}\left( \frac{1}{\sqrt{2\bar E j^2}} \right)\,,
\eeq
and
\beq
\chi^{\rm tail}(\bar E, j)= \frac{1}{\nu} \frac{\partial}{\partial j} \int \frac{dr}{p_r^{(0)}} H^{\rm  tail }(r,p_r^{(0)},j)\,.
\eeq
We thereby see that the tail correction to the scattering function derives, via a $j$-gradient, from the following  \lq\lq potential" $W^{\rm tail}(\bar E,j)$ 
\begin{eqnarray}
\label{W_tail}
W^{\rm tail}(\bar E,j)&=&\int \frac{dr}{p_r^{(0)}} H^{\rm  tail }(r,p_r^{(0)},j)\nonumber\\
&=&\left\llbracket\int dt  H^{\rm  tail } \right\rrbracket_{(\bar E, j)}\,.
\end{eqnarray}
In the second expression we have used the property that $dr/p_r^{(0)}=dt$ along the Newtonian Hamiltonian flow, as well as the notation introduced above for signifying
that a quantity is computed along a specified flow line of some given zeroth-order Hamiltonian flow.
The resulting formula
\beq
\label{eq_e_delta_chi_hamilt}
\chi^{\rm tail}(\bar E,j) =  \frac{1}{\nu} \frac{\partial}{\partial j} W^{\rm tail}(\bar E,j)\,,
\eeq
can also be viewed as a hyperbolic analogue of the Delaunay averaging procedure.

The Delaunay approach to ellipticlike motions shows that when one is dealing with a perturbed Hamiltonian of the type
\beq
H({\mathbf q},{\mathbf p})=H_0({\mathbf q},{\mathbf p})+\epsilon H_1({\mathbf q},{\mathbf p})
\eeq
one can, modulo a canonical transformation of order $\epsilon$, eliminate all oscillatory terms in $H_1({\mathbf q},{\mathbf p})$, thereby replacing $H_1({\mathbf q},{\mathbf p})$ by its time average. Similarly here, the result \eqref{W_tail} shows that one can eliminate from $H^{\rm tail}({\mathbf q},{\mathbf p})$ any total time derivative contribution (vanishing at infinite separation).  

Let us exhibit the explicit form of the potential $W^{\rm tail}(\bar E,j)$.
To this aim, we start form  Eq. \eqref{H_tail_I3_I3}.  To simplify the writing, let us introduce the notation (where $t'=t+\tau$)
\beq
F(t,t')\equiv \frac15   I_{ij}^{(3)}(t)  I_{ij}^{(3)}(t')\,,
\eeq
for the \lq\lq time-split" gravitational wave flux. [Note that, as said above, we use here units where $M$ and $G$ are set to unity.
In particular, the quadrupole moment we use corresponds to $I_{ij}^{\rm phys}/(GM)^2$.]

With this notation the potential $W^{\rm tail}$, Eq. \eqref{W_tail}, reads
\begin{eqnarray}
\label{int_H_tail_loc_time}
W^{\rm tail}(\bar E,j)=-  \int dt \, {\rm Pf}_{2s/c}\int_{-\infty}^\infty \frac{dt' }{|t'-t|} F(t,t')\,.\nonumber\\
\end{eqnarray}
An exact expression for  $F(t,t')$ along the Newtonian motion (valid both for elliptic and for hyperbolic motions,
without any approximation in eccentricity) is
\begin{eqnarray}
\label{flux_fun}
F(t,t')&=&\frac{4\nu^2 }{15j^{10}}(1+e\cos \phi' )^2\times \nonumber\\
&\times & (1+e\cos \phi )^2\left( F_0+F_1 e +F_{2} e^2\right)\,,
\end{eqnarray}
where
\begin{eqnarray}
F_0(\phi,\phi')&=&  24\cos(2 \phi-2\phi' )\nonumber\\
F_1(\phi,\phi')&=& 9\cos(2\phi-3\phi')+15\cos(-2\phi'+\phi)\nonumber\\
&+& 9\cos(3\phi-2\phi')+15\cos(2\phi-\phi') \nonumber\\
F_{2}(\phi,\phi')&=& -\frac14 \cos(\phi+\phi')+\frac{45}{8}\cos(3\phi-\phi')\nonumber\\
&+&\frac{45}{8}\cos(-3\phi'+\phi)+\frac{27}{8}\cos(3\phi-3\phi')\nonumber\\
&+&\frac{77}{8}\cos(\phi-\phi') \,.
\end{eqnarray}
Here we have re-expressed the phase space variables ${\mathbf q}$ and ${\mathbf p}$ entering $I_{ij}^{(3)}$ 
entirely in terms of the two azimuthal angles 
\beq
\phi =\phi(t)\,,\qquad \phi'=\phi(t')\,,
\eeq
using the Keplerian orbit relations
\beq
r=\frac{j^2}{1+e\cos \phi}\,,\qquad r'=\frac{j^2}{1+e\cos \phi'}\,.
\eeq
In order to explicitly compute the two time integrals entering the expression \eqref{int_H_tail_loc_time} for $W^{\rm tail}$ one further needs the explicit time dependence of $\phi$ and $\phi'$.
In the presently considered case of hyperbolic motions, this has to be done via the hyperbolic Kepler equation, namely
\beq
\label{Kepler}
\bar n \, t = e \sinh \bar u -\bar u\,,
\eeq
where 
\beq
\bar n = \frac{1}{\bar a^{3/2}}\,,\qquad \bar a= \frac{1}{2\bar E}=-a\,.
\eeq
We recall also that the parameter ${\sf p}$ of the considered conic satisfies
\beq
{\sf p}=a(1-e^2)=\bar a (e^2-1)=j^2\,,
\eeq
so that we have also
\beq
e^2= 1 + 2\bar E \, j^2 = 1 + v_{\infty}^2 \,  j^2\,.
\eeq
Then, given the solution $\bar u(\bar n t,e)$ of Kepler's equation \eqref{Kepler},
one has
\beq \label{phiu}
\phi(t)=2 \, {\rm arctan}\left[\sqrt{\frac{e+1}{e-1}}\tanh \left(\frac{\bar u}{2} \right)  \right]\,.
\eeq

A different, but equivalent, expression for $W^{\rm tail}(\bar E,j)$ can be obtained by working in the Fourier domain.
Our notation for the Fourier-transform $\hat I_{ij}(\omega)$ of the Newtonian quadrupole moment of the system is
\begin{eqnarray}
I_{ij}(t)&=&\int \frac{d\omega}{2\pi} \hat I_{ij}(\omega) e^{-i\omega t}\nonumber\\
\hat I_{ij}(\omega)&=&\int dt I_{ij}(t) e^{i\omega t}\,.
\end{eqnarray}
Inserting these Fourier representations (together with their  \lq\lq primed" counterparts and the notation $T_s\equiv 2s/c$) yields
\begin{widetext}
\begin{eqnarray}
W^{\rm tail}(\bar E,j)&=& \int dt \, H_{\rm tail}= -\frac{1}{5}\, \int dt \, {\rm Pf}_{T_s}\int \frac{dt'}{|t-t'|} I_{ij}^{(3)}(t)I_{ij}^{(3)}(t')\nonumber\\
&=& -\frac{1}{5} \int dt {\rm Pf}_{T_s}\int\frac{dt'}{|t-t'|} \int \frac{d\omega}{2\pi} \frac{d\omega'}{2\pi} (-i\omega)^3 (-i\omega')^3  \hat I_{ij}(\omega) \hat I_{ij}(\omega') e^{-i\omega t}e^{-i \omega'  t'} \nonumber\\
&=& -\frac{1}{5} {\rm Pf}_{T_s}\int \frac{d\tau}{|\tau|}  \int \frac{d\omega}{2\pi} \frac{d\omega'}{2\pi} (-i\omega)^3 (-i\omega')^3  \hat I_{ij}(\omega) \hat I_{ij}(\omega') \int dt e^{-i\omega t}e^{-i \omega'  (t+\tau)}
\end{eqnarray}
where  we came back to the notation $t'=t+\tau$. Using the fact that  integral $\int dt e^{-i(\omega+\omega')t}=2\pi \delta (\omega+\omega')$ we get
\begin{eqnarray}
W^{\rm tail}(\bar E,j)
&=& -\frac{1}{5}  \int \frac{d\omega}{2\pi}   \omega^6 \hat I_{ij}(\omega) \hat I_{ij}(-\omega)  {\rm Pf}_{T_s}\int \frac{d\tau}{|\tau|} e^{i\omega  \tau} \,.
\end{eqnarray}
\end{widetext}
The partie finie integral entering the latter expression is  (see Eq. (5.8) in Ref. \cite{Damour:2014jta}; with $\gamma=\gamma_{\rm Euler}$)
\begin{eqnarray}
{\rm Pf}_{T_s}\int_{-\infty}^\infty \frac{d\tau}{|\tau|} e^{i\omega  \tau} 
&=& -2 \ln (|\omega| {T_s} e^\gamma)\,,
\end{eqnarray}
so that, using the relation $\hat I_{ij}^*(-\omega)=\hat I_{ij}(\omega)$, we find the final Fourier-domain formula\footnote{Though we often set $c$ to one, we leave it here as a reminder that $s$ is a length scale, while $T_s\equiv 2s/c$ is a time scale.} 
\beq
\label{W_fourier}
W^{\rm tail}(\bar E,j) 
= \frac{2}{5}  \int_{-\infty}^\infty \frac{d\omega}{2\pi}   \omega^6 |\hat I_{ij}(\omega)|^2  \ln \left(|\omega| \frac{2s}{c} e^\gamma\right)\,.
\eeq
This remarkably compact formula is very simply related to the frequency-domain version of  the total gravitational wave energy emitted by the entire hyperbolic motion 
which reads
\beq
\label{I_fourier_flux}
\Delta E_{\rm GW}
= \frac{1}{5}  \int_{-\infty}^\infty \frac{d\omega}{2\pi}   \omega^6 |\hat I_{ij}(\omega)|^2 \,.
\eeq
More precisely, the only difference consists in  inserting the factor $2 \ln \left(|\omega| \frac{2s}{c} e^\gamma\right)$ in the frequency-decomposition of $\Delta E_{\rm GW}$.
As a consequence,  the $s$-dependence of $W^{\rm tail}$ is entirely contained in the term
\beq
W^{\rm tail}(\bar E,j)\Bigg|_{\ln s}=2 \, \Delta E_{\rm GW}(\bar E,j) \, \ln s\,,
\eeq
where we recall the known result for gravitational-wave energy loss along hyperbolic orbits \cite{turner} (see also Ref. \cite{Blanchet:1989cu} for its 1PN generalization)
\begin{eqnarray}\label{EGW}
\frac{\Delta E_{\rm GW}}{M c^2}&=&\frac{2\nu^2 }{15j^7}\left[ (37e^4 +292e^2 +96) \arccos\left( -\frac{1}{e} \right)\right. \nonumber\\
&&\left. +\frac13 \sqrt{e^2-1}(673 e^2 +602) \right]\,.
\end{eqnarray}
Here, we came back provisorily to usual physical units.

To turn the result \eqref{W_fourier} into an explicit function of $\bar E$ and $j$ we need to insert in it the Fourier-transform of the quadrupole moment of the system computed along a hyperbolic Newtonian orbit. The steps for doing so are the following.

First we note that the $\mu$-rescaled quadrupole moment reads
\beq
\frac{1}{\mu}  I_{ij}=\begin{pmatrix}
\frac23 x^2 -\frac13 y^2 & xy & 0\\
xy & \frac23 y^2 -\frac13 x^2 & 0\\
0 & 0 &-\frac13 x^2 -\frac13 y^2\\
\end{pmatrix}
\,.
\eeq
Second, we rewrite the expressions of the Fourier-transforms of $x^2$, $y^2$ and $xy$, with the notation
$(x^2)_\omega=\int dt x^2(t) e^{i\omega t}$, etc., as (here we assume $\omega>0$, and we go back to $GM$-scaled units)
\begin{eqnarray}
\label{rel_x2y2xx_final}
(x^2)_\omega &=&-\frac{2\pi}{u^2v_\infty^7}\left[ \left(-u \, e+\frac{u}{e}+i \right)H_{\frac{iu}{e}}^{(1)}(iu)\right. \nonumber\\
&&\left. +u (e^2-1)H_{\frac{iu}{e}+1}^{(1)}(iu)\right] \nonumber\\
(y^2)_\omega &=& -\frac{2\pi (e^2-1)}{e^2 v_\infty^7 u^2}\left[\left(\frac{u}{e}+i  \right) H_{\frac{iu}{e}}^{(1)}(iu) \right. \nonumber\\
&& \left. - u  H_{\frac{iu}{e}+1}^{(1)}(iu) \right]\nonumber\\
(x y)_\omega &=& \frac{2\pi \sqrt{e^2-1}}{u^2v_\infty^7}\left[ \left(-u \, e+\frac{u}{e}+i \right)H_{\frac{iu}{e}}^{(1)}(iu)\right. \nonumber\\
&& \left. -ie H_{\frac{iu}{e}+1}^{(1)}(iu)\right]\,.
\end{eqnarray}
Here we used the notation
\beq \label{unotation}
u=\frac{\omega}{\bar n} e>0\,,\quad e=\sqrt{1+2\bar E j^2}=\sqrt{1+v_\infty^2 j^2}
\eeq
together with the standard notation $H_{\nu}^{(1)}(z)$ for the Hankel functions of the first type.
[The so-introduced notation $u$ for the argument of the Hankel functions
should not be confused with the previous notation $u=1/r$.]
See Appendix \ref{Hankel} for the details of the derivation of the expressions \eqref{rel_x2y2xx_final}.

\section{Explicit computation of $W^{\rm (tail)} = \int dt \, H^{\rm (tail)}$ and $\chi^{\rm (tail)}$
 in the large eccentricity limit}

In this section we will compute, at leading order (LO) in $1/e$, the potential $W^{\rm (tail)} = \int dt \, H^{\rm (tail)}$
(whose $j$-derivative yields the tail contribution $\chi^{\rm (tail)}$ to the scattering function, as per Eq. \eqref{eq_e_delta_chi_hamilt}).
We will give two independent calculations of $W^{\rm (tail)}$. One, in the time domain, i.e. using Eq. \eqref{int_H_tail_loc_time},
and the other, in the Fourier domain, i.e. using Eq. \eqref{W_fourier}. In addition, we shall also check, in Appendix \ref{RungeLenz}, the 
correctness of the value of $\chi^{\rm (tail)}$  obtained by differentiating $W^{\rm (tail)}$, by means of a direct dynamical computation involving the evolution
of the Laplace-Lagrange-Runge-Lenz vector.

\subsection{Time-domain computation of  $W^{\rm (tail)} = \int dt \, H^{\rm (tail)}$}

In the following, we shall consider hyperbolic motions in the vicinity of an infinite eccentricity (corresponding
to straight-line, uniform motion), and expand the scattering potential $W^{\rm (tail)}$ in powers of $1/e$.
A way to set up such a large-$e$-expansion is to introduce the eccentricity-rescaled mean motion
\beq
\tilde n \equiv \frac{\bar n}{e} = \frac{(2\bar E)^{3/2}}{\sqrt{1+2\bar E j^2}} \,.
\eeq
In terms of $\tilde n$, the Kepler equation \eqref{Kepler} reads
\beq
\label{Kepler2}
\tilde n \, t =  \sinh \bar u - \frac1e \bar u\,.
\eeq
The solution $\bar u (\tilde n t)$ of the rescaled Kepler equation \eqref{Kepler2} can then be expanded in powers of $1/e$
(keeping fixed the rescaled time $\tilde t \equiv \tilde n t$) :
\beq
\bar u (\tilde n t , e)= \arcsinh \tilde n t + \frac1e {\bar u}_1 (\tilde n t) + \frac1{e^2} {\bar u}_2 (\tilde n t) + \cdots
\eeq
Inserting this expansion in $\phi(\bar u)$, Eq. \eqref{phiu}, then yields the $1/e$ expansion of  $\phi(\tilde n t)$:
\beq \label{phieexp}
\phi(t) = \phi_0(t)+\frac{1}{e}\phi_{1}(t) +\frac{1}{e^2}\phi_{2}(t) + \cdots
\eeq
The LO terms read 
\beq
\phi_0(t) ={\rm arctan}(\tilde n t) \,,
\eeq
\beq
 \phi_1(t)=\frac{\tilde n t}{\sqrt{1+\tilde n^2 t^2}}+\frac{\arcsinh(\tilde n t)}{1+\tilde n^2 t^2} \,.
\eeq
Note than an alternative way to derive the expansion \eqref{phieexp} is to start from Kepler's area law
\beq \label{arealaw}
r^2 \frac{d\phi}{dt} = j \,.
\eeq
Inserting the polar equation
\beq
r(t)=\frac{j^2}{1+e\cos(\phi)}
\eeq
in the area law \eqref{arealaw}  yields the following differential equation for $\phi(t)$
\beq
\frac{d\phi}{dt}=\frac{1}{j^3}(1+e\cos\phi)^2\,,
\eeq
which can be integrated term by term in the $1/e$ expansion.

Inserting the expansion \eqref{phieexp} in the exact expression \eqref{flux_fun} of the time-split gravitational-wave flux $F(t,t')$
then yields a large-$e$-expansion of $F(t,t')$ of the form (denoting $t' \equiv t+ \tau$)
\beq
F(t,t + \tau) = e^6  f_6(t,\tau) +  e^5 f_5(t,\tau) + \cdots
\eeq
where
\beq
f_6(t,\tau)=\frac{4 \nu^2}{15j^{10}} \cos^2\phi_0' \cos^2\phi_0\,\, F_2(\phi_0,\phi_0')\,,
\eeq
where $\phi_0 \equiv \phi_0(t) = \arctan (\tilde n t)$, and $\phi_0' \equiv \phi_0(t+\tau) = \arctan (\tilde n (t+\tau))$.
Inserting these solutions leads to the compact expression
\beq
f_6(t,\tau)=\frac{4 \nu^2}{15 j^{10}}\frac{{\sf f}_6(\tilde t,\tilde \tau)}{[1+(\tilde t+\tilde \tau)^2]^{5/2}(1+{\tilde t}^2)^{5/2}}\,,
\eeq
where we used (on the right-hand-side) the rescaled time variables ${\tilde t} \equiv \tilde n t$, ${\tilde \tau} \equiv \tilde n \tau$, and where
\begin{eqnarray} \label{f6}
{\sf f}_6(  {\tilde t}, {\tilde \tau})&=& 24+50 {\tilde t}^2+56 {\tilde t}^3 {\tilde \tau}+39 {\tilde t}^2{\tilde \tau}^2+6 {\tilde t}^5{\tilde \tau}+6 {\tilde t}^4{\tilde \tau}^2\nonumber\\
&+&11 {\tilde t}{\tilde \tau}^3+2 {\tilde t}^3{\tilde \tau}^3-12 {\tilde \tau}^2+50{\tilde \tau} {\tilde t}+28 {\tilde t}^4+2 {\tilde t}^6 \,. \nonumber\\
\end{eqnarray}

Let us now show how, using the above large-$e$ expansion, we can, in principle, compute the large-$e$ expansions of
both the time-localized value of the tail Hamiltonian  $H_{\rm (loc)}^{\rm (tail)}$, Eq. \eqref{H_tail_loc_time}, and
the corresponding scattering potential $W^{\rm (tail)}= \int dt H_{\rm (loc)}^{\rm (tail)}$. Here we shall explicitly compute
only the LO term in the large-$e$ expansions of $H_{\rm (loc)}^{\rm (tail)}$ and  $W^{\rm (tail)}$.
In principle, our approach gives a way to compute all the terms in the large-$e$ expansion, though the higher-order
terms involve integrals that seem difficult to compute explicitly. [We had a first look at the problem and found, however,
that the large-$e$ expansion of $H_{\rm (loc)}^{\rm (tail)}(t)$ can be evaluated in terms of polylogarithm functions.]

The evaluation of the tail Hamiltonian  $H_{\rm (loc)}^{\rm (tail)}$, Eq. \eqref{H_tail_loc_time}, crucially involves the partie finie
of an integral over $  {\tilde \tau} = \tilde n \tau$. [In view of the rescaling by $\tilde n$, the corresponding regularization
scale becomes $\tilde T \equiv 2 \tilde n s/c$.] Namely,
\beq
\frac1{e^6} H^{\rm (loc) \, LO }_{\rm (tail)}({\tilde t})=-{\rm Pf}_{\tilde T}\int \frac{d \tilde \tau}{| {\tilde \tau}|}f_6(\tilde t, {\tilde \tau}) \,.
\eeq
Using the explicit expression of $f_6(\tilde t, {\tilde \tau})$, Eq. \eqref{f6}, we found that it was straightforward to compute
the regularized integral, with the LO result
\begin{eqnarray}
\frac1{e^6} H^{\rm (loc) \, LO}_{\rm (tail)}({\tilde t}) &=& -\frac{8 \nu^2}{15 j^{10}}\frac{2({\tilde t}^2+12)}{(1+{\tilde t}^2)^3}\ln \left(\frac{2}{{\tilde T}}(1+{\tilde t}^2)\right)\nonumber\\
&&+ \frac{8}{15 j^{10}}\frac{7{\tilde t}^2+36}{(1+{\tilde t}^2)^3}\,.
\end{eqnarray}
Finally, we can compute the scattering potential $W^{\rm (tail)}= (1/\tilde n) \int d{\tilde t }H_{\rm (loc)}^{\rm (tail)}$ (also at the leading order)
\begin{eqnarray} \label{WLO1}
{ W}^{\rm (tail) \, LO} 
&=&\frac{2}{15}\frac{e^6}{\tilde n} \frac{\pi}{j^{10}}\nu^2 \left[ 100+37 \ln \left( \frac{\tilde T}{8}\right)\right]\nonumber\\
&\approx&\frac{2}{15}\pi \nu^2 \frac{v_\infty^4}{j^3}\left[ 100+37 \ln \left( \frac{sv_\infty^2}{4j}\right)\right] \,.
\end{eqnarray}
In the last expression, we have used the LO large-$e$ approximations
\beq
e\approx v_\infty j \, , \quad \tilde n \approx \frac{v_\infty^2}{j} \,.
\eeq

\subsection{Fourier-domain computation of $W^{\rm tail}=\int dt H_{\rm (tail)}$} 

It is useful to complement the time-domain computation of $W^{\rm tail}=\int dt H_{\rm (tail)}$ presented in the previous subsection
by its frequency-domain counterpart, obtained by considering the large-eccentricity expansion of our general result \eqref{W_fourier}.

Let us start from the Fourier transform of the quadrupole moment, Eqs. \eqref{rel_x2y2xx_final}. [See Appendix \ref{Hankel} for details of 
its derivation.]  Here, we will limit our discussion to the  leading order in the large-eccentricity limit, where the general expressions simplify. 
We consider the $e\to \infty$ limit of Eqs. \eqref{rel_x2y2xx_final} keeping fixed the previously defined, Eq. \eqref{unotation},
argument of the Hankel functions:
\beq
u= \frac{\omega}{\bar n} e = \frac{\omega}{\tilde n}  \,.
\eeq
Indeed, our time-domain large-$e$ expansion above has shown that the rescaled time variable adapted to this limit
is $ \tilde n t$, so that $\tilde n$ yields the characteristic frequency of the large-$e$ hyperbolic motion. One then expects
that the Fourier spectrum will be essentially localized around the characteristic frequency $\tilde n$, i.e. for values of 
$u= \omega/{\tilde n}$ of order one. We then see on Eqs. \eqref{rel_x2y2xx_final} that the limit $e\to \infty$ simplifies
the order of the Hankel functions to the values 0 or 1. We note in passing that the latter Hankel functions are related via
\beq
H_0^{(1)}{}'(q)=-H_1^{(1)}(q)\,.
\eeq
The LO large-$e$ limit of Eqs. \eqref{rel_x2y2xx_final} yields
\beq
\bar n \approx v_\infty^3\,,\qquad \bar a \approx \frac{1}{v_\infty^2}\,,
\eeq
and
\begin{eqnarray}
(x^2)_\omega &=& -i \pi \frac{2e^2}{v_\infty^7}\frac{1}{q}H^{(1)}_1(q) \nonumber\\
(y^2)_\omega &=& i \pi \frac{2e^2}{q^2v_\infty^7}(q H^{(1)}_1(q)+ H^{(1)}_0(q))     \nonumber\\
(x y)_\omega &=& - i \pi \frac{2e^2}{q^2v_\infty^7} (q H^{(1)}_0(q)-H^{(1)}_1(q) )  \,,
\end{eqnarray}
where we denoted $q= i u = i \omega/{\tilde n}$. The latter expressions involve Hankel functions evaluated at purely imaginary
arguments. Let us now recall that $H^{(1)}_\nu(iz)$ is simply related to the modified Bessel function, $K_\nu$,  at argument $z$.
 (see \cite{AS}, Eq. (9.6.4)). When the order $\nu$ is zero or one, this relation reads
\beq
K_0(x)=i \frac{\pi}{2}H^{(1)}_{0}(i x)\,,\qquad K_1(x)=-\frac{\pi}{2}H^{(1)}_{1}(i x)\,.
\eeq
This yields
\begin{eqnarray} \label{quadsplash}
(x^2)_\omega &=&    \frac{4e^2}{v_\infty^7}\frac{1}{ u}  K_1(u)\nonumber\\
(y^2)_\omega &=&     -  \frac{4e^2}{u^2 v_\infty^7}(  u K_1(u)+ K_0(u))\nonumber\\
(x y)_\omega &=&   i  \frac{4e^2}{u^2 v_\infty^7} (  u K_0(u) +K_1(u) )\,.
\end{eqnarray}
Inserting these expressions in Eq. \eqref{inertia_tot} leads to
\beq \label{splash1}
u^6 |\hat I_{ij}(u)|^2 = \frac{32 e^4}{v_\infty^{14}}{\mathcal F}(u)\,,
\eeq
where we defined
\begin{eqnarray}\label{splash2}
{\mathcal F}(u) &\equiv& \left( \frac{u^2}{3}+u^4 \right)K_0^2(u) +3u^3K_0(u)K_1(u)\nonumber\\
&&+(u^2+u^4)K_1^2(u) \,.
\end{eqnarray}
Equivalently, this means that the energy flux per frequency interval reads
\beq \label{splash3}
\frac{d\omega}{2\pi} \omega^6 |\hat I_{ij}(\omega)|^2 =\frac{32 v_\infty^{4} }{2\pi j^3} {\mathcal F}(u)du\,.
\eeq
Note that the integrand has to be doubled if one integrates on  $\omega$ (or $u$) only from $0$ to $+\infty$, instead of $-\infty$ to $+\infty$.

The results \eqref{quadsplash},  \eqref{splash1}, \eqref{splash2}, \eqref{splash3}, are equivalent to the classic results of 
Ruffini and Wheeler (see Ref. \cite{Rees:1974iy}, pag. 127, and Ref. \cite{rr-wheeler}) on  the \lq\lq splash gravitational radiation" from a particle in fast hyperbolic motion. To check the equivalence with their results it is useful to introduce the impact parameter (both in physical
units and in its rescaled version)
\beq
b^{\rm phys}= \frac{L}{\mu v_\infty}; \quad  \quad b = \frac{b^{\rm phys}}{GM}= \frac{j}{v_\infty}\,.
\eeq
The argument $u$ of the Bessel functions above is then (consistently with the notation of Refs. \cite{Rees:1974iy,rr-wheeler})
\beq
u = \frac{\omega}{\tilde n} = \frac{\omega b}{v_\infty} \,,
\eeq
while the gravitational-wave (one-sided; $f = \omega^{\rm phys}/(2 \pi) >0$) spectrum reads
\beq
\frac{d\Delta E}{df}=\frac{64 G}{5}  \left( \frac{G \mu  M}{b^{\rm phys}}\right)^2\, {\mathcal F}(u) \,.
\eeq
The integrated gravitational-wave flux, i.e. the total energy emitted by the splash radiation is
\beq
\Delta E_{\rm splash} = \frac{64}{5}  \frac{v_\infty}{2\pi b^3}(\mu  M)^2 \int_0^\infty {\mathcal F}(u) du\, .
\eeq
One finds that
\beq
\int_0^\infty {\mathcal F}(u) du=\frac{37\pi^2}{96} \,,
\eeq
so that
\beq
\frac{\Delta E_{\rm splash}}{M}=\frac{37}{15} \pi \nu^2 \frac{v_\infty}{b^3}= \frac{37}{15} \pi \nu^2 \frac{v_\infty^4}{j^3}\,.
\eeq
Note that this result agrees with the large-$e$ limit of Eq. \eqref{EGW} (remembering $ e\approx v_\infty j$).

After this check, let us come back to our main purpose, namely the Fourier-domain computation of the scattering potential $W^{\rm tail}$.
It is given by the following logarithmically-modified version of the total emitted gravitational-wave energy (taking into account the factor $2$
linked to the one-sidedness of the integral)
\begin{eqnarray}
W^{\rm tail} &=&  2   \int_0^\infty df \frac{d\Delta E}{df} \ln \left(|\omega| \frac{2s}{c}e^\gamma  \right)\\
&=&
  \frac{64}{5 \pi}  \nu^2 \frac{v_\infty^4}{j^3}   \int_0^\infty   du  {\mathcal F}(u) \ln \left(|\omega| \frac{2s}{c}e^\gamma  \right)\,,\nonumber
\end{eqnarray}
where
\beq
\omega=\frac{v_\infty u }{b} = \frac{v_\infty^2 u }{j}\,
\eeq
so that the logarithm reads $\ln(\alpha u)$ with $\alpha = 2s v_\infty^2 e^\gamma/j$.

Now, we found that
\begin{eqnarray}
\int_0^\infty {\mathcal F}(u)  \ln(\alpha u) du &=&\pi^2 \left[\frac{25}{24}+\frac{37}{96}\ln\left(\frac{\alpha}{8e^\gamma}\right) \right]\nonumber\\
&=&\frac{\pi^2}{96}
\left[100 + 37\ln\left(\frac{\alpha}{8e^\gamma}\right) \right]
\,.
\end{eqnarray}
Substituting the value of $\alpha=2s v_\infty^2 e^\gamma/j$ then gives 
\begin{eqnarray}\label{WLO2}
\frac{W^{\rm tail}}{M}  &=&\frac{1}{M}\int dt  H_{\rm (tail)}
\nonumber\\
&=& \frac{2}{15}\nu^2  \pi \frac{v_\infty^4}{j^3}
\left[100  + 37\ln\left(\frac{sv_\infty^2}{4j}\right) \right] \,,
\end{eqnarray}
in agreement with the result of our time-domain computation Eq. \eqref{WLO1}.

\subsection{Computation of the tail contribution to scattering in the large-eccentricity limit}

Having confirmed the LO computation of the scattering potential, Eqs. \eqref{WLO1}, \eqref{WLO2}, 
we can now differentiate  $W^{\rm tail}(\bar E,j)$ with respect to $j$ (at fixed $\bar E$, i.e. at fixed $v_\infty$)
to get the corresponding LO tail contribution to the scattering angle.

Putting back the energy scale $M c^2$ in Eq. \eqref{eq_e_delta_chi_hamilt}, but keeping the integral over the dimensionless
time $t= c^3 T/GM$, so that
\beq
\chi_{\rm tail}^{\rm LO}=\frac{1}{\nu}\partial_j \left( \frac{1}{M c^2}\int dt  H_{\rm (tail)}\right)= \frac{1}{\nu}\partial_j \left( \frac{W^{\rm tail}}{M c^2}\right)
\eeq
we finally get
\begin{eqnarray} \label{chitailLO}
 \chi_{\rm tail}^{\rm LO} &=&   -\frac{2}{15}\pi  \nu  \frac{v_\infty^4}{j^4} \left[337+111\ln\left(\frac{{ s} v_\infty^2}{4j}\right)\right]\nonumber\\
&=& -\frac{2}{ 5}\pi  \nu  \frac{v_\infty^4}{j^4} \left[\frac{337}{3}+37\ln\left(\frac{ { s} v_\infty^2}{4j}\right)\right]
\,,
\end{eqnarray}
where we recall that $  s \equiv c^2 s^{\rm phys}/GM$ is a dimensionless regularization scale defining the nearzone-farzone separation.

As a further check on our result \eqref{chitailLO}, we have shown that it agrees with a direct, dynamical computation based on
the secular evolution of the Newtonian-conserved Laplace-Lagrange-Runge-Lenz, eccentricity vector under the action of the 
nonlocal, tail  4PN-level force ${\pmb{ \mathcal F}}^i_{\rm (nonloc)}$, Eq. \eqref{F_non_loc}.  The reader will find this
alternative calculation in Appendix \ref{RungeLenz}.

\section{Summing the local and nonlocal (tail) contributions to $\chi^{\rm 4PN}$ in the large-$j$ limit}

We have separately computed above (using an expansion in powers of $1/(j v_\infty )$ when necessary) both the local and nonlocal (tail) contributions to the scattering function  $\chi(\bar E, j)$, with 4PN accuracy.
Our results at the 1PN, 2PN and 3PN levels (see Eqs. \eqref{PNexpchi}) were given as exact, closed-form expressions. By contrast, 
the 4PN-level value $\chi^{\rm 4PN}$ of the function\footnote{Note that the notion of ``$n$-PN-level contribution" to $\chi$ delicately
depends on the choice of the energy variable used as argument for the function $\chi(\bar E, j)$. The use of another energy measure,
such as $E- M c^2$ or ${\mathcal E}_{\rm eff}$  would lead to a complete reshuffling of each individual $n$-PN-level contribution,
while leaving invariant only the sum over all PN levels.} was obtained as a sum of a closed-form local contribution (modulo the
logarithmic term $\mathcal I_{\chi}$), and of a nonlocal one (given as a large-eccentricity expansion). 
Let us here combine the two separate (local and nonlocal) 4PN-level
contributions to $\chi(\bar E, j)$. 
Replacing $\bar E$ by $\frac12 v_\infty^2$ at 4PN we find

\begin{eqnarray}
\label{chi_loc_j_gg12}
\frac{\chi^{jv_\infty  \gg 1}_{\rm loc}}{2}\bigg|_{\rm 4PN}&=& \frac{v_\infty^6}{j^2}\pi\left[  \frac{27}{64}\nu^2+\frac{9}{64} \nu-\frac{15}{32} \nu^3  \right]\nonumber\\
&+& \frac{v_\infty^5}{j^3}\left[  -8\nu^3-\frac{91}{5}\nu+34\nu^2 \right]\nonumber\\
&+& \frac{v_\infty^4}{j^4}\pi \left[-\frac{225}{32} \nu^3+\left(\frac{4827}{64} -\frac{369}{512}\pi^2\right)\nu^2 \right. \nonumber\\
&+& \left(-\frac{91537}{960}  +\frac{33601}{16384}\pi^2 +\frac{37}{5} \ln\left(\frac{v_\infty s}{2j}\right)\right)\nu \nonumber\\
&+& \left.
\frac{3465}{128}   \right]\nonumber\\
&+& O\left(\frac{v_\infty^3}{j^5}\right)\nonumber\\
\frac{\chi^{jv_\infty   \gg 1}_{\rm tail} }{2}\bigg|_{\rm 4PN}&=&   -\frac{1}{ 5}\pi  \nu  \frac{v_\infty^4}{j^4} \left[\frac{337}{3}+37\ln\left(\frac{sv_\infty^2}{4j}\right)\right]\nonumber\\
&+ & O\left(\frac{v_\infty^3}{j^5}\right)\,.
\end{eqnarray}
Combining these two pieces only modifies the term $O(\nu)$ in ${v_\infty^4}/{j^4}$ and the final 4PN-level  result  is the following
\begin{eqnarray}
&&\frac{ \chi^{jv_\infty   \gg 1}_{\rm loc }+ \chi^{\j \gg 1}_{\rm tail}}{2}\bigg|_{\rm 4PN}=
\frac{v_\infty^6}{j^2}\pi\left[  \frac{27}{64}\nu^2+\frac{9}{64} \nu-\frac{15}{32} \nu^3  \right]\nonumber\\
&&\quad +\frac{v_\infty^5}{j^3}\left[  -8\nu^3-\frac{91}{5}\nu+34\nu^2 \right]\nonumber\\
&&\quad + \frac{v_\infty^4}{j^4}\pi\left[-\frac{225}{32}\nu^3+\left(\frac{4827}{64} -\frac{369}{512}\pi^2\right)\nu^2 \right. \nonumber\\
&&\quad + \left(\frac{33601}{16384}\pi^2+\frac{37}{5} \ln\left(\frac{2}{v_\infty}\right)-\frac{22621}{192}\right)\nu \nonumber\\
&& \quad +\left. 
\frac{3465}{128}    \right]
+O\left(\frac{v_\infty^3}{j^5}\right)\,.
\end{eqnarray}
It is important to note that the arbitrary scale $s$ has cancelled between the two contributions (as it should),
but that it has left a ``large-logarithm" contribution $\propto  \ln\left(\frac{2}{v_\infty}\right)= \ln\left(\frac{2 \, c}{v_\infty^{\rm phys}}\right)$.

\section{Including NNLO linear-in-spin contributions to the scattering angle}

To complete our 4PN-accurate computation of the orbital (i.e. non-spinning) contribution to the scattering, let us now tackle
the additional contributions due to spin effects. For simplicity, we restrict ourselves to the (non-precessing) case of parallel spins,
and we work only linearly in the spins. 

One of the additional advantages of working within the EOB formalism is that it is rather straightforward to include the linear-in-spin
contributions at NNLO to the scattering angle. Indeed, it is enough to complete the local calculation of Section \ref{sec:local}
by including in the PN expansion of the function $p_r(\bar E,j)$ the linear-in-spin contributions.
This is done by including in the  EOB  energy conservation law both  the orbital, and the  spin-orbit, terms. In other
words, we start with  the the spin-dependent conserved energy
\begin{eqnarray} 
\label{Efspin}
\hat {\mathcal E}_{\rm eff}&=&\sqrt{A(1+j^2 u^2 +A\bar D p_r^2+\hat Q)}\nonumber\\
&&+\frac{u^3}{M^2} j [g_S S+g_{S_*}S_*] \,.
\end{eqnarray}
The two (phase-space-dependent) dimensionless  gyrogravitomagnetic ratios $g_{S}$ and $g_{S*}$ are  known (in the PN-expanded sense) at the next-to-next-to-leading-order  (NNLO) level \cite{Nagar:2011fx,Barausse:2011ys}
\begin{eqnarray}
\label{gssstar_PN}
g_S^{\rm PN}(u,p_r,p_\phi) &=&2 +\eta^2 \left[-\frac58 \nu u-\frac{27}{8}\nu p_r^2   \right]\nonumber\\
&+& \eta^4 \left[\nu \left(-\frac{51}{4}u^2-\frac{21}{2}up_r^2+\frac58 p_r^4 \right)\right. \nonumber\\
&+& \left. \nu^2 \left( -\frac18 u^2 +\frac{23}{8}u p_r^2+\frac{35}{8}p_r^4 \right) \right]\nonumber\\
&+& O(\eta^6)\nonumber\\
g_{S*}^{\rm PN}(u,p_r,p_\phi) &=& \frac32 \nonumber\\
&+& \eta^2 \left[-\frac98 u -\frac{15}{8}p_r^2  +\nu \left(-\frac34 u-\frac94 p_r^2  \right)\right]\nonumber\\
&+& \eta^4 \left[-\frac{27}{16}u^2+\frac{69}{16}up_r^2+\frac{35}{16}p_r^4 \right. \nonumber\\
&+&  \nu \left(-\frac{39}{4}u^2-\frac{9}{4}u p_r^2+\frac52 p_r^4 \right)\nonumber\\
&+&\left. \nu^2 \left( -\frac{3}{16}u^2 +\frac{57}{16}u p_r^2+\frac{45}{16}p_r^4 \right) \right]\nonumber\\
&+& O(\eta^6)\,. 
\end{eqnarray}
The values of $g_{S}$ and $g_{S*}$ cited above have been expressed in the Damour-Jaranowski-Schaefer (DJS) spin gauge \cite{Damour:2007nc,Damour:2008qf}, which is defined so that these quantities do not actually depend on $p_\phi$. Let us note in passing that 
recent gravitational self-force computations have extended (for the terms  linear in the symmetric mass ratio $\nu$),  the
knowledge of $g_{S}$ and $g_{S*}$ to a  high PN level \cite{Bini:2014ica,Bini:2015xua}. We will, however, not make use
of this (partial)  knowledge here.

Passing, as above,  to the energy variable $\bar E=\frac{1}{2}(\hat {\mathcal E}_{\rm eff}^2-1)$ and
solving (in a PN sense) the energy conservation law \eqref{Efspin} for $p_r=p_r^{(0)}+\eta p_r^{(1)}+ \eta^2 p_r^{(2)} + \eta^3 p_r^{(3)}+\ldots $ (where $\eta=1/c$) , leads, in addition to the terms involving  even powers of $\eta$ that we used in Section \ref{sec:local} 
to the following additional, $\eta$-odd contributions:
\begin{widetext}
\begin{eqnarray}
 p_r^{(1)} &=& 0 \nonumber\\
 p_r^{(3)} &=&   -\frac{j u^3}{2 M^2 p_r^{(0)}}(4 S+3 S_*) \nonumber\\
 p_r^{(5)} &=&   \frac{j u^3}{8 M^2 p_r^{(0)}{}^3}[A^{(5,S)}S + A^{(5,S_*)}S_*]\nonumber\\
 p_r^{(7)} &=&  -\frac{u^3 j}{16 M^2 p_r^{(0)}{}^5} [A^{(7,S)}S + A^{(7,S_*)}S_*]
\end{eqnarray}
where
\begin{eqnarray}
A^{(5,S)}&=& (-108\bar E j^2 u^2+27 j^4 u^4+118 u^2-113 j^2 u^3+226 u\bar E+108\bar E^2)\nu   \nonumber\\
&&  -64 u^2+48 j^2 u^3-96 u\bar E+16\bar E j^2 u^2-32 \bar E^2 \nonumber\\
A^{(5,S_*)}&=&  (-78 j^2 u^3+84 u^2+72\bar E^2+18 j^4 u^4-72\bar E j^2 u^2+156 u\bar E)\nu \nonumber\\
&&   +66 u\bar E+30 u^2+15 j^4 u^4-48\bar E j^2 u^2-33 j^2 u^3+36\bar E^2 \nonumber\\
A^{(7,S)}&=& (-7816 u^2 \bar E^2-1680 u^4 j^4 \bar E^2-4848 u \bar E^3+7816 u^4 \bar E j^2+560 u^6 j^6 \bar E-1480 u^4-1120 \bar E^4+606 u^7 j^6\nonumber\\
&& +2240 \bar E^3 j^2 u^2-1954 u^6 j^4+7272 u^3 \bar E^2 j^2-3636 u^5 j^4 \bar E-70 u^8 j^8+2784 j^2 u^5-5568 u^3 \bar E) \nu^2\nonumber\\
&&+(4632 u \bar E^3-6064 u^3 \bar E^2 j^2+13504 u^2 \bar E^2-12404 u^4 \bar E j^2-358 u^7 j^6+4992 u^4+84 u^4 j^4 \bar E^2-10 u^8 j^8\nonumber\\
&&-328 \bar E^3 j^2 u^2+26 u^6 j^6 \bar E+14136 u^3 \bar E+272 \bar E^4-6636 j^2 u^5+2590 u^5 j^4 \bar E+2826 u^6 j^4) \nu\nonumber\\
&& -512 u^4+256 u^3 \bar E^2 j^2-1280 u^3 \bar E+960 u^4 \bar E j^2-128 u \bar E^3-960 u^2 \bar E^2+16 u^4 j^4 \bar E^2+640 j^2 u^5+64 \bar E^4\nonumber\\
&& -96 u^5 j^4 \bar E-240 u^6 j^4-64 \bar E^3 j^2 u^2\nonumber\\
A^{(7,S_*)}&=& (-3336 u \bar E^3+5004 u^3 \bar E^2 j^2-5676 u^2 \bar E^2+1440 \bar E^3 j^2 u^2-1080 u^4 j^4 \bar E^2-1164 u^4-720 \bar E^4+417 u^7 j^6\nonumber\\
&& +360 u^6 j^6 \bar E-1419 u^6 j^4-2502 u^5 j^4 \bar E+5676 u^4 \bar E j^2-45 u^8 j^8+2112 j^2 u^5-4224 u^3 \bar E) \nu^2\nonumber\\
&&+(368 u \bar E^3+2336 u^4-324 u^5 j^4 \bar E+5792 u^3 \bar E-744 u^4 j^4 \bar E^2+4176 u^2 \bar E^2+104 u^7 j^6+284 u^6 j^6 \bar E\nonumber\\
&& -352 \bar E^4-40 u^8 j^8-2608 j^2 u^5+848 \bar E^3 j^2 u^2+48 u^3 \bar E^2 j^2-3432 u^4 \bar E j^2+672 u^6 j^4) \nu\nonumber\\
&& +196 u^4+1548 u^3 \bar E^2 j^2+712 \bar E^3 j^2 u^2+100 j^2 u^5+199 u^7 j^6+250 u^6 j^6 \bar E+280 u^3 \bar E-35 u^8 j^8\nonumber\\
&& -648 u^4 j^4 \bar E^2-272 \bar E^4-1008 u^5 j^4 \bar E+936 u^4 \bar E j^2-656 u \bar E^3-300 u^2\bar E^2-393 u^6 j^4\,.
\end{eqnarray}
and
\beq
 p_r^{(0)}=\sqrt{2\bar E-j^2 u^2+2 u}\,.
\eeq
We compute then
\beq
 U(\bar E, j, \nu; u)=-\frac{1}{u^2}\partial_j p_r
\eeq
and finally we obtain  the (half) of the scattering angle
\beq
\frac{1}{2}(\chi+\pi)={\rm Pf}\int_0^{u_{\rm max}}U(\bar E, j, \nu; u) du
\eeq
with, as usual, $u_{\rm max}$ computed at the Newtonian approximation 
\beq
u_{\rm max}=\frac{1}{j^2}[1+\sqrt{1+2\bar E j^2}]\,.
\eeq
The result is the following 
\begin{eqnarray}
\frac{\chi^{\rm LO}}{2} &=& -\frac{2}{j^3}\left[ B(\alpha) +\frac{(1+2\alpha^2)}{2\alpha (1+\alpha^2)} \right] \left(2S+\frac32 S_{*}  \right)\nonumber\\
\frac{\chi^{\rm NLO}}{2} &=& \frac{1}{j^5}\left[ C^{\rm B,NLO}B(\alpha) +C_0^{\rm NLO} \right] \nonumber\\
\frac{\chi^{\rm NNLO}}{2} &=& \frac{1}{j^7}\left[ C^{\rm B,NNLO}B(\alpha) +C_0^{\rm NNLO} \right]  \,,
\end{eqnarray}
where
\begin{eqnarray}
C^{\rm B,NLO}&=& \frac{1}{2\alpha^2}[C^{\rm B,NLO,S}S + C^{\rm B,NLO,S_*}S_*]\nonumber\\
C_0^{\rm NLO}&=&  \frac{1}{8\alpha^3 (1+\alpha^2)^2}[C_0^{\rm NLO,S}S+C_0^{\rm NLO,S_*}S_* ]
\end{eqnarray}
with
\begin{eqnarray}
C^{\rm B,NLO,S}&=& (21\alpha^2+8)\nu-168\alpha^2-24\nonumber\\
C^{\rm B,NLO,S_*}&=& (1+3\alpha^2)6 \nu -105 \alpha^2 -12 \nonumber\\
C_0^{\rm NLO,S}&=&  (1+\alpha^2)(84 \alpha^4 +88 \alpha^2 +9)\nu -672\alpha^6 -1216 \alpha^4 -520\alpha^2 -8 \nonumber\\
C_0^{\rm NLO,S_*}&=& 6(1+\alpha^2)(12\alpha^4+12\alpha^2+1)\nu-420\alpha^6-748\alpha^4-305\alpha^2-1  \,.
\end{eqnarray}
Similarly
\begin{eqnarray}
C^{\rm B,NNLO}&=& -\frac{1}{8M^2 \alpha^4}[C^{\rm B,NLO,S}S + C^{\rm B,NLO,S_*}S_*]\nonumber\\
C_0^{\rm NNLO}&=&  -\frac{1}{16 \alpha^5  (1+\alpha^2)^3M^2}[C_0^{\rm NLO,S}S+C_0^{\rm NLO,S_*}S_* ]
\end{eqnarray}
with
\begin{eqnarray}
C^{\rm B,NNLO,S}&=&  (150\alpha^4+168\alpha^2+32)\nu^2+(-5265\alpha^4-2376\alpha^2-94)\nu+11880\alpha^4+3360\alpha^2+36  \nonumber\\
C^{\rm B,NNLO,S_*}&=& 3 [ (45\alpha^4+48\alpha^2+8)\nu^2+(-1320\alpha^4-544\alpha^2-14)\nu+2310\alpha^4+560\alpha^2+1)] \nonumber\\
C_0^{\rm NNLO,S}&=&  (300\alpha^{10}+1136\alpha^8+1620\alpha^6+1046\alpha^4+276\alpha^2+14)\nu^2\nonumber\\
&&+(-10530\alpha^{10}-32832\alpha^8-36026\alpha^6-15767\alpha^4-2050\alpha^2-7)\nu\nonumber\\
&& +23760\alpha^{10}+70080\alpha^8+70264\alpha^6+25836\alpha^4+2016\alpha^2-4 \nonumber\\
C_0^{\rm NNLO,S_*}&=&  (270\alpha^{10}+1008\alpha^8+1410\alpha^6+885\alpha^4+222\alpha^2+9)\nu^2\nonumber\\
&&+(-7920\alpha^{10}-24384\alpha^8-26212\alpha^6-11022\alpha^4-1272\alpha^2+2)\nu\nonumber\\
&&+13860\alpha^{10}+40320\alpha^8+39458\alpha^6+13743\alpha^4+840\alpha^2 -1) \,.
\end{eqnarray}
The part of the background scattering angle $\lim_{\nu\to 0}\chi$ proportional to $S$ has been independently checked  by studying geodesic  motion in the Kerr spacetime, at linear order in the rotational parameter of the black hole.

\end{widetext}

\section{Determination of the variation of  $W^{\rm tail}(e,j)$ (and $H^{\rm tail}$) with the eccentricity in the full hyperbolic domain $1\leq e\leq +\infty$}

We have explicitly computed above the LO approximation to the large-eccentricity expansion of the tail
scattering potential $W^{\rm tail}$, and we have also indicated that, in principle, modulo the tackling
of more complicated integrals, our large-$e$ expansion can be continued to higher orders in $1/e$.
Here we wish to complete this discussion by briefly considering the global behavior of  $W^{\rm tail}(e,j)$, considered as a function of $e$ and $j$ 
(rather than $\bar E= v_\infty^2/2$ and $j$), when, for a fixed value of $j$, the eccentricity varies
over the complete range of hyperbolic motions, i.e. from $e=\infty$ down to the parabolic case where $e=1$.

First, we note that our general Fourier-domain result Eq. \eqref{W_fourier} suggests to write the value of $W^{\rm tail} = \int dt H^{\rm tail}$
in the form
\beq \label{Wej1}
W^{\rm tail}(e,j) =  2 \, \Delta E_{\rm GW}(e,j) \, \ln \left(\omega_c(e ,j) \frac{s}{c} \right)
\eeq
where the total gravitational-wave energy loss $\Delta E_{\rm GW}(e,j)$ is given by Eq. \eqref{EGW} \cite{turner},
and where $\omega_c(e,j)$ is some characteristic frequency of the gravitational-wave spectrum.

Our LO computation above (see \eqref{WLO1} with $v_\infty \approx e/j$) has shown that, when $e \gg 1$, the characteristic frequency $\omega_c(e,j)$ (or rather
its logarithm) is asymptotically equivalent to
\beq \label{Wej2}
\ln \omega_c(e,j)  \approx \ln \left( \frac{e^2}{4j^3}\right) + \frac{100}{37} \, , \,{\rm when} \, e\to \infty.
\eeq

Actually, when considering what is the characteristic orbital frequency of hyperbolic motions as $e$ varies between $\infty$ and $1$,
one can easily see that it will parametrically scale as $e^2/j^3$ over the full range of hyperbolic motions. In particular, it must
be of order $1/j^3$ for parabolic motions ($e=1$), because this is the only frequency scale for such motions.
In addition, if we had considered the  gravitational-wave spectrum in usual (say CGS) physical units, as the scale $s$ enters
the tail logarithm in the form of a {\it time scale} $s^{\rm phys}/c$ (as recalled in Eq. \eqref{Wej1}), the physical-units version of
the characteristic frequency  $\omega_c(e,j)$ must indeed have the dimension of $[{\rm time}]^{-1}$. Dimensional analysis
then shows that, modulo a function of the dimensionless eccentricity $e$, $\omega_c(e,j)$ must scale proportionally
to the inverse cube of the angular momentum. In other words, coming back to our scaled units, we conclude that
the product $\omega_c(e,j) j^3$ can only be a function of the (dimensionless) eccentricity $e$.
In conclusion, we can parametrize the behavior of $W^{\rm tail}(e,j)$ over the full range of hyperbolic motions
by writing Eq. \eqref{Wej1} together with an expression for $\omega_c(e,j)$ of the form
\beq \label{logomc}
\ln \omega_c(e,j)  \equiv \ln \left( \frac{e^2}{4j^3}\right) + C(\frac1e) \,,
\eeq
where the dimensionless contribution $C(\frac1e)$ is only a function of $e$.

The issue of controlling the behavior of $W^{\rm tail}(e,j)$ over the full range of hyperbolic motions is then
reduced to controlling the variation of the (dimensionless) function $C(\e)$ as $\e \equiv \frac1e$ increases from 0 to 1.
We already know from Eq. \eqref{Wej2} that
\beq \label{C0}
C(0) = \frac{100}{37}\,.
\eeq
We have studied the variation of $C(\e)$ over the full
interval $0 \leq \e \equiv \frac1e \leq1$ by two different approaches: (i) by an analytical study of the limit $\e=1$ (parabolic motion);
and (ii) by a numerical study of the Fourier-domain expression of $W^{\rm tail}(e,j)$ for a sample of intermediate
values $0 < \e < 1$. Let us here only briefly summarize our results.

In dimensionless units parabolic orbits ($\e\equiv1/e=1$) can be parametrized as follows [see \cite{LL-mech}, pag. 75, Ex. 1]
\begin{eqnarray}
x&=&\frac{p}{2}(1-\eta^2)\,,\qquad y = p\eta\,,\qquad r=\frac{p}{2}(1+\eta^2)\nonumber\\ 
t&=&p^{3/2}\frac{\eta}{2}\left(1+\frac{\eta^2}{3}  \right)\,,
\end{eqnarray}
where $p=j^2$ and $\eta\in (-\infty, \infty)$. 
Comparing  with the polar representation of the parabolic orbit,
\beq
r=\frac{j^2}{1+\cos \phi}=\frac{j^2}{2 \cos^2 \frac{\phi}{2}}\,,
\eeq
we find that
\beq
\eta = \tan \frac{\phi}{2}\,, \qquad \phi=2 \arctan \eta\,.
\eeq
The time-split flux function $F(t, t')$, \eqref{flux_fun}, then becomes
\begin{eqnarray}
F(\eta, \eta') &=& \frac{512}{15 j^{10}(1+\eta^2)^5 (1+\eta'{}^2)^5}P(\eta,\eta') \,.
\end{eqnarray}
where
\begin{eqnarray}
P(\eta,\eta')&=& (\eta^5+8\eta^3-11\eta)\eta'{}^5\nonumber\\
&+& (8\eta^5+76\eta^3-112\eta)\eta'{}^3\nonumber\\   
&+&(-60+300\eta^2)\eta'{}^2\nonumber\\
&+&(-112\eta^3+169\eta-11\eta^5)\eta'\nonumber\\
&+& 12-60\eta^2\,.
\end{eqnarray}
We then transform the partie-finie integral over $t'$ we are interested in, namely
\beq
{\rm Pf}_{2 s/c}\, \int \frac{dt'}{|t-t'|}F(t,t')
\eeq
in an integral over $\eta'$, using $dt'=\frac{j^3}{2}(1+\eta'{}^2)d \eta'$, $|t-t'|=\frac16 |\eta-\eta'|(\eta^2+\eta \eta' +\eta'{}^2+3)$, and 
also taking care of correspondingly changing the regularization time scale into $T_\eta = 2 s/[c(dt'/d\eta')_{\eta}]= 4 s/(c j^3(1+\eta^2))$:
\beq
 {\rm Pf}_{T_\eta} \, \int \frac{d\eta'}{|\eta-\eta'|}  \frac{F(\eta, \eta')}{(\eta^2+\eta \eta' +\eta'{}^2+3)}\,.
\eeq
We then find  
\begin{widetext}
\begin{eqnarray}
{\rm Pf}_{2s}\, \int \frac{dt'}{|t-t'|}F(t,t')&=& \frac{1}{j^{10}}  \frac{512}{5}\left\{
\frac13 \frac{\eta^2+12}{(1+\eta^2)^6}\ln \left( \frac{3}{{T_{\eta}}^2}(1+\eta^2) \right)
-\frac{\eta^6+30\eta^4+230 \eta^2+288}{(1+\eta^2)^6 (\eta^2+4)^2} \right. \nonumber\\
&& \left.
+\frac{2}{\sqrt{3}}  \frac{(\eta^6 +4\eta^4-66 \eta^2+28)}{(1+\eta^2)^6 (\eta^2+4)^{5/2}}{\rm arctan}\left(\frac{\sqrt{3}\eta}{\sqrt{\eta^2+4}}\right)
\right\}\nonumber\\
&\equiv & \frac{1}{j^{10}}{\mathcal F}(\eta)\,.
\end{eqnarray}
\end{widetext}

We can then compute $W_{\rm tail} = \int dt H_{\rm tail}$ by integrating over $t$ (using $dt=\frac{j^3}{2}(1+\eta{}^2)d \eta$):

\begin{eqnarray}
W_{\rm tail}(e=1,j)&=&
-\int dt {\rm Pf}_{2s}\, \int \frac{dt'}{|t-t'|}F(t,t')\nonumber\\
&=&-\int \frac{j^3}{2}(1+\eta^2) \frac{1}{j^{10}}{\mathcal F}(\eta) d\eta \nonumber\\
&=&
\frac{340}{3j^7}\pi \frac{M\nu^2}{c^5} \left[\frac{416}{85}-\frac12 \ln 3 +\ln \left(\frac{s }{4j^3}\right) \right]
 \,.\nonumber\\
\end{eqnarray}
Factoring
\beq
2 \Delta E_{\rm BS}(e=1,j)=\frac{340}{3}\frac{\pi}{c^5} M \frac{\nu^2}{j^7}
\eeq
we then get a result which can be written as in Eq. \eqref{Wej1}, with $\ln \omega_c$ of the form of Eq. \eqref{logomc} with
the dimensionless additional constant having the value
\beq \label{C1}
C(1) = \frac{416}{85}-\frac12 \ln 3\approx  4.344811502\,.
\eeq
By comparison, we recall the value  $C(0)=100/37=2.\overline{702}$, obtained above in Eq. \eqref{C0}.

To complete our analytical determination of the function $C(\e)$ at the two extreme values $\e=0$, Eq. \eqref{C0}, and $\e=1$, 
Eq. \eqref{C1},  of hyperbolic motions, we have also computed numerical estimates  of $C(\e)$ at a sample of intermediate values
of the inverse eccentricity: $0< \e=1/e<1$. We found useful to compute the function  $W_{\rm tail}(e,j)$ by rewriting our Fourier-domain
formula Eq. \eqref{W_fourier} in terms of the rescaled frequency variable $v$, defined as
\beq
v\equiv \frac{\omega}{\omega_*}\,, \, \,{\rm with} \quad  \omega_*\equiv\frac{e^2}{j^3}.
\eeq
Indeed, we have already mentioned that $\omega_*$ measures, for all values of $e \geq1 $, the characteristic frequency of gravitational wave emission.
In terms of $v$ our Fourier-domain integral reads
\begin{eqnarray}
\label{Wv}
&&W_{\rm tail}(e,j) = \frac{2}{5\pi}  \frac{e^{14}}{j^{21}}\times \nonumber\\
&&\quad \int_0^\infty d v\, v^6 |I_{ij}(v;e,j)|^2 \ln \left(2s e^\gamma \frac{e^2}{ j^3}  v \right)\,.
\end{eqnarray}

Using \eqref{Wv}, we computed the numerical values of the additional constant $C(\e)$ in Eq. \eqref{logomc} for a sample
of intermediate values of $\e$ in the hyperbolic-motion interval $0<\e<1$. They are listed in Table I. 
\[
\begin{array}{|c|c|c|}
\hline
 \e =1/e& C(\e) & c(\e)\cr
\hline
0 & 2.702702703 & 0\cr
10^{-5}&  2.702727104 & 0.1485955134\times 10^{-4}\cr
1/52345 & 2.702749308 & 0.2838118889\times 10^{-4}\cr
10^{-4} & 2.702946704 & 0.1485900326\times 10^{-3}\cr
10^{-3} & 2.705141653 & 0.1485254815\times 10^{-2}\cr
10^{-1} & 2.935443483 & 0.1417328621\cr
1/8 &     2.990307450 & 0.1751435393\cr
1/6.2 &   3.067754516 & 0.2223067151\cr
1/4.33  & 3.209535882 & 0.3086477457\cr
1/3.62  & 3.297190578 & 0.3620270931\cr
1/2.73 &  3.461705299 & 0.4622121241\cr
1/2.21 &  3.608040700 & 0.5513264392\cr
1/2.13 &  3.635709539 & 0.5681760165\cr
1/1.5  &  3.931344289 & 0.7482096112\cr
1/1.412 &  3.988242935 & 0.7828593530 \cr
1/1.29 &  4.076465716 & 0.8365846489\cr
1/1.13 &  4.212464303 & 0.9194041229\cr
1/1.0345 &4.307440862 & 0.9772422875\cr
1/1.0243 & 4.318308127& 0.9838601590\cr
1 & 4.344811502       & 1 \cr
\hline
\end{array}
\]

We have completed Table I by listing  in the last column the values of the normalized version, $c(\e)$, of $C(\e)$,  defined as
\begin{eqnarray} \label{defc}
c(\e)&=&\frac{C(\e)-C(0)}{C(1)-C(0)} \,, \nonumber\\ 
C(\e)&=& C(0) + \left[C(1)-C(0) \right] c(\e)\,.
\end{eqnarray}
Note that $c(\e)$ varies between 0 and 1 as $\e$ varies between 0 and 1. [In the following we assume that $C(0)$ and $C(1)$ 
take their exact analytical values \eqref{C0} and \eqref{C1}.]
We have explored several possible simple, analytic fits for $c(\e)$. For instance, the cubic polynomial 
\beq 
\label{cubicfit}
c_{\rm cubic}(\e)=1.4730\e -0.6318\e^2 + 0.1588\e^3\,,
\eeq
agrees with the numerical data of Table 1within $6\cdot 10^{-4}$ (maximum difference). A more
accurate representation (maximum difference  $5\cdot 10^{-6}$) is given by the following Pad\'e approximant
\beq\label{Padefit}
c_{\rm Pade'}(\e)= \e \frac{0.018265  \e^2-0.715430 \e+1.485639}{1-0.211528\e^2}\,.
\eeq

The data points for $C(\e)$ given in Table I are plotted in Fig. 1, and compared there with the simple cubic fit \eqref{cubicfit}
\begin{figure}
\includegraphics[scale=0.35]{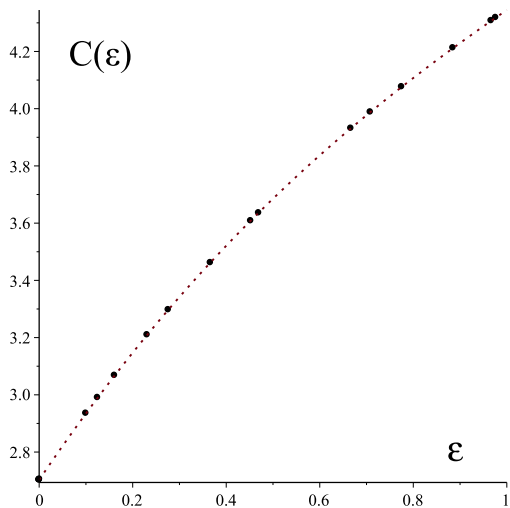}
\caption{\label{fig_varie_ecc} The behavior, as a function of $\e=1/e$, of the additional constant $C(\e=1/e)$ in Eq. \eqref{logomc} is plotted using both the data points of  table I, and the simple cubic fit \eqref{cubicfit} for the normalized version \eqref{defc} of $C(\e)$.}
\end{figure}

By inserting in Eqs. \eqref{Wej1}, \eqref{Wej2}, \eqref{logomc},  the results on $C(\e)$ we have just given, we obtain a global
representation of the scattering potential $W^{\rm tail}(e,j) =\int dt H^{\rm tail}$ (and therefore of $\chi^{\rm tail}$) 
over the full range of hyperbolic motions. [Note that, in order to
derive $\chi^{\rm tail}$ from $W^{\rm tail}(e,j)$, we have to replace $e$ as a function of $\bar E$ and $j$ (namely 
$e=\sqrt{1+ 2 \bar E j^2}$ before differentiating with respect to $j$ (at fixed energy).]

Let us go one step further, and deduce from our results a corresponding global description of the non-integrated tail Hamiltonian $H^{\rm tail}(q,p)$ over the full range of hyperbolic motions. [Here, we denote by $(q,p)$ some canonical phase-space coordinates, say $(r,p_r, \phi, p_\phi)$.] There are several ways of constructing such a (time-localized) tail Hamiltonian. Let us start by remarking that the factor 
$\Delta E_{\rm GW}(e,j)$ in Eq. \eqref{Wej1} is the time integral of the instantaneous gravitational-wave flux
\beq
 {\mathcal F}_E(q,p)=\frac{8\nu^2}{5r^4}\left(4 p^2-\frac{11}{3} p_r^2\right) \,.
\eeq
As a consequence, we can define (using one of the global representations of $C(1/e)$ discussed above)
\beq \label{HtailC}
H^{\rm tail, C}(q,p) \equiv + 2 \, {\mathcal F}_E(q,p) \, \left[\ln \left(\frac{e^2 s}{4 \, c \, p_\phi^3}\right) + C(\frac1e)  \right]_{e=e(q,p)}\,.
\eeq
This local function of $(r,p_r, \phi, p_\phi)$ decreases as $1/r^4$ when the binary separation increases, and its time-integral 
is equal to $W^{\rm tail}(e,j)$. Thereby, $H^{\rm tail, C}(q,p)$ provides a time-localized global description of the tail
contribution to the Hamiltonian over the full range of hyperbolic motions. It differs from our original time-localized Hamiltonian
Eq. \eqref{H_tail_loc_time} by some ($O(1/c^8)$) canonical transformation (corresponding to adding a total time derivative).

When considering the vicinity of  $1/e=0$, we have shown that, modulo an additional canonical transformation, one can
replace the Hamiltonian \eqref{HtailC} by the following more explicit function of positions and momenta
\begin{eqnarray}
\frac{H'_{\rm tail}}{M}
&=& + 2 \, {\mathcal F}_E(q,p) \ln\left(\frac{j s}{c\, r^2}\right)\nonumber\\ 
&&+\frac{8 }{15 r^4}\nu^2 \left( 36 {\bf p}^2 -29  p_r^2\right) +O\left(\frac{1}{e} \right)\,.
\end{eqnarray}
The latter,  tail Hamiltonian is the large-eccentricity counterpart of the (time-localized) small-eccentricity tail Hamiltonian introduced
in Ref. \cite{Damour:2015isa} which reads, at order $O(e^0)+O(e^2)$  (after using a canonical transformation with generating function proportional to
$p_r/r^3$  to trade the terms $\propto 1/r^5$ by terms $\propto ({\bf p}^2 - 4 p_r^2)/r^4$)
\begin{eqnarray}
\frac{H_{\rm tail}}{M}&=&+2 {\mathcal F}_E(q,p) \, \ln \left( \frac{s}{c \, r^{3/2}}\right) \nonumber\\
&+& \frac{8 }{15 r^4}\nu^2 \, \left( \alpha_{\rm ell}\, {\bf p}^2 +\beta_{\rm ell}\, p_r^2\right)  +O\left(e^4 \right)\,,
\end{eqnarray}
with 
\begin{eqnarray}
\alpha_{\rm ell} &=&24 \ln (4e^\gamma)\nonumber\\
\beta_{\rm ell} &=&-198-22\gamma +\frac{2187}{4}\ln 3 -598\ln 2 \,.
\end{eqnarray}

\section{Concluding remarks}

Using the recently computed 4PN-level EOB Hamiltonian \cite{Damour:2015isa} (involving both local and nonlocal parts), we
have computed here for the first time the corresponding 4PN-accurate (conservative\footnote{A way to deal with the effect of radiation-reaction
on $\chi$ has been given in Ref. \cite{Bini:2012ji}}) scattering angle $\chi$ of hyperboliclike encounters of
(comparable-mass)  two-body systems. The scattering angle was previously known only up to the 2PN level \cite{Bini:2012ji}.
While our computation of the local part of $\chi$ makes use of rather well established  PN techniques (simplified by the use of the EOB formalism), our computation of the tail part is based on novel techniques. In addition, we have completed our computation of $\chi$ by providing the
contributions linear in the spins of the two bodies, at the next-to-next-to-leading PN order.

Both as a check on our computation, and as a way to provide several different viewpoints on the scattering of binary systems, we
have implemented several different methods for computing the nonlocal, tail contribution to $\chi$, which is the most
subtle part of our calculations. 

First, we introduced a generalization of the time-localization technique introduced in Refs. \cite{Damour:2015isa,Damour:2016abl}.
Indeed, the (Delaunaylike) time-localization technique of  Refs. \cite{Damour:2015isa,Damour:2016abl} was limited to the
case of ellipticlike motions. Here, we have shown how to generalize this idea to the case of hyperboliclike motions. Our final result
is that the tail part $\chi_{\rm tail}$ of the scattering angle can be derived via the $j$-gradient of a tail potential: 
$W^{\rm tail}(\bar E,j)=\int H^{\rm tail} dt$. We then gave two different methods for computing the scattering potential 
$W^{\rm tail}(\bar E,j)$. One method works directly in the time-domain, while a second method works in the frequency-domain.
The final result of the latter method, given in Eq. \eqref{W_fourier}, illuminates the physics of $W^{\rm tail}(\bar E,j)$ by
showing that it is a logarithm-weighted avatar of the frequency-decomposition of the gravitational-wave emission of binary
systems.

In addition, we showed (in Appendix \ref{RungeLenz}) how to compute $\chi_{\rm tail}$ 
by considering the evolution (during the hyperbolic encounter) 
of the Laplace-Lagrange-Runge-Lenz eccentricity vector ${\mathbf A}$.  This additional computation provides
a nice check of our large-eccentricity generalization of the time-localization technique.

Besides providing general formulas for $\chi_{\rm tail}$, we have also shown that the functional dependence of the
scattering potential $W_{\rm tail}(e,j)$ on the eccentricity $e$ and the angular momentum $j$ could be described,
through Eqs. \eqref{Wej1}, \eqref{logomc} (where $\Delta E_{\rm GW}(e,j)$ is the integrated gravitational wave luminosity,
Eq. \eqref{EGW}), by means of a function of the sole eccentricity, namely the function $C(1/e)$
entering Eq. \eqref{logomc}. We have analytically computed the value of $C(\e)$ (where $\e\equiv 1/e$)
for $\e=0$ (large eccentricity limit) and for $\e=1$ (parabolic motions). In addition, we have numerically estimated
the value of $C(\e)$ for a sample of intermediate inverse-eccentricity values $0<\e<1$, and provided some simple analytic fits
allowing one to describe $C(\e)$ in the full interval $0\leq\e\leq 1$.  This allowed us to analytically describe both $W_{\rm tail}$,
and a specific time-localized version of the tail Hamiltonian $H_{\rm tail}(q,p)$, over the full range of hyperbolic motions (see 
Eq. \eqref{HtailC}).
We have also shown how
our results for $\chi_{\rm tail}$ in the large-eccentricity limit are related to the classic results of
Ruffini and Wheeler on the ``gravitational splash radiation" \cite{Rees:1974iy,rr-wheeler}.

In addition to the main results summarized above, we have also provided: (i) a sketch of an alternative method for computing
the local part $\chi_{\rm loc}$ of the scattering angle (see Appendix \ref{alternativemethod}); and (ii) the explicit form of
a local Hamiltonian yielding the large-eccentricity limit of the scattering.

Summarizing, this work is a contribution to the  programme \cite{Bini:2012ji,Damour:2014afa,Damour:2016gwp} of extending the EOB formalism beyond the inspiral-plunge-merger
regime of binary systems to the regime of high-energy hyperbolic encounters of binary systems.

\appendix

\section{Computing $\hat I_{ij}(\omega)/\mu$ in terms of Hankel's functions} \label{Hankel}

We give here some details of the derivation of the Fourier transform $\hat I_{ij}(\omega)$ of the quadrupole moment along binary
hyperbolic motions, as given in Eqs. \eqref{rel_x2y2xx_final}.

The nonvanishing components of $\hat I_{ij}(\omega)$ are\footnote{Here we work, without explicitly indicating it,
with the rescaled center-of-mass frame quadrupole moment  $I_{ij}/\mu$, and we use $GM$-rescaled space and time units.}
\begin{eqnarray}
\hat I_{xx}(\omega)&=& \left(\frac23 x^2-\frac13 y^2\right)_\omega \nonumber\\
\hat I_{yy}(\omega)&=& \left(\frac23y^2-\frac13 x^2  \right)_\omega \nonumber\\
\hat I_{xy}(\omega)&=&  (x y)_\omega\nonumber\\
\hat I_{zz}(\omega)&=&- \left(\frac13 x^2+\frac13 y^2\right)_\omega \,,
\end{eqnarray}
so that the square of $\hat I_{ij}(\omega)$   reads
\begin{eqnarray}
\label{inertia_tot}
|\hat I_{ij}(\omega)|^2&=& \frac{2}{3}\left[|(x^2)_\omega|^2+|(y^2)_\omega|^2\right] +2|(x y)_\omega|^2\nonumber\\
&& -\frac13 \left[(x^2)_\omega (y^2)_{-\omega}+(x^2)_{-\omega} (y^2)_{\omega}\right] \,.
\end{eqnarray}

We are interested here in (Newtonian level) hyperbolic motions. This is conveniently parametrized as
\begin{eqnarray}
x&=&-a (\cosh \bar u-e)\nonumber\\ 
y&=&-a \sqrt{e^2-1}\sinh \bar u\nonumber\\ 
\bar n t &=& e \sinh \bar u -\bar u\,,
\end{eqnarray}
with $a=(-2\bar E)^{-1}$ and $\bar n=(-a)^{-3/2}=v_\infty^3$. [Note that, when discussing hyperbolic motion (with $\bar E >0$), it is
sometimes convenient to replace $a$ by $\bar a \equiv -a = (+2\bar E)^{-1} >0$.]

The Fourier transform of the quadrupole tensor is computed by replacing the time integration by an integration over $\bar u$
(later denoted as $\xi$):
\begin{eqnarray}
(x^2)_\omega &=& \int dt e^{i \omega t} x^2(t)\nonumber\\
&=&a^2 \int \frac{dt}{d\bar u} (\cosh \bar u-e)^2 e^{i \frac{\omega}{\bar n}  (e\sinh \bar u-\bar u)} d \bar u\nonumber\\
&=&
\frac{a^2}{\bar n}\int d\xi (e\cosh \xi -1)(\cosh \xi-e)^2 e^{i \frac{\omega}{\bar n}  (e\sinh \xi-\xi)}
\nonumber\\
(y^2)_\omega &=& \int dt e^{i \omega t} y^2(t)\nonumber\\
&=&a^2 (e^2-1) \int \frac{dt}{d\bar u} \sinh^2  \bar u e^{i \frac{\omega}{\bar n}  (e\sinh \bar u-\bar u)} d \bar u
\nonumber\\
&=&
\frac{a^2 (e^2-1)}{\bar n}\int d\xi (e\cosh \xi -1)\sinh^2 \xi  e^{i \frac{\omega}{\bar n}  (e\sinh \xi-\xi)}\nonumber\\
(x y)_\omega &=& \int dt e^{i \omega t} x(t)y (t)\nonumber\\
&=&a^2\sqrt{e^2-1} \int \frac{dt}{d\bar u}   (\cosh \bar u-e) \sinh   \bar u \, e^{i \frac{\omega}{\bar n}(e\sinh \bar u-\bar u)}   d \bar u\nonumber\\
&=&
\frac{a^2 \sqrt{e^2-1}}{\bar n}\int d\xi (e\cosh \xi -1)(\cosh \xi-e) \times \nonumber\\
&& \times \sinh   \xi  \, e^{i \frac{\omega}{\bar n}  (e\sinh \xi-\xi)} \,,
\end{eqnarray}
from which we see that $(x^2)_\omega=[(x^2)_{-\omega}]^*$.

The above integrals can be expressed in terms of  Hankel's functions of the first kind $H^{(1)}$  (see Eq. 9.1.25 in Ref. \cite{AS}) 
\beq
\int_{-\infty}^\infty e^{q \sinh \xi - p \xi} \, d\xi = i \pi H^{(1)}_{p}(q)\,
 \,. 
\eeq

In our case, the argument $q$ and the order $p$ are
\beq
 q=i e \frac{\omega}{\bar n} \,,\qquad p=i \frac{\omega}{\bar n}=\frac{q}{e} \,.
\eeq

A direct computation (based on decomposing in exponentials of $\xi$ the factors in front of 
$e^{i \frac{\omega}{\bar n}  (e\sinh \bar u-\bar u)}$) gives
\begin{eqnarray}
\label{generali}
(x^2)_\omega &=&i \pi \frac{a^2 }{2\bar n}\left[e\left(\frac{11}{4}+e^2  \right)[H^{(1)}_{p-1}(q)+H^{(1)}_{p+1}(q)]\right.\nonumber\\
&&-\left(e^2+\frac12 \right)[H^{(1)}_{p-2}(q)+H^{(1)}_{p+2}(q)]\nonumber\\
&&+\frac{e}{4}[H^{(1)}_{p-3}(q)+H^{(1)}_{p+3}(q)] \nonumber\\
&&\left. -(1+4e^2)H^{(1)}_{p}(q)\right]\nonumber\\
(y^2)_\omega &=&i \pi \frac{a^2(e^2-1)}{4\bar n}\left[\frac{e }{2}[H^{(1)}_{p-3}(q)+H^{(1)}_{p+3}(q)]\right.\nonumber\\ 
&&- [H^{(1)}_{p-2}(q)+H^{(1)}_{p+2}(q)]\nonumber\\
&&\left. +\frac{e  }{2 }[H^{(1)}_{p-1}(q)+H^{(1)}_{p+1}(q)]  +2H^{(1)}_{p}(q) \right]\nonumber\\
(x y)_\omega &=& i \pi \frac{a^2 \sqrt{e^2-1} }{4\bar n}\left[-\frac{  e}{2} [H^{(1)}_{p-3}(q)-H^{(1)}_{p+3}(q)]\right. \nonumber\\
&& +(e^2+1)  [H^{(1)}_{p-2}(q)-H^{(1)}_{p+2}(q)]\nonumber\\
&&\left. -\frac{5   e }{2}  [H^{(1)}_{p-1}(q)-H^{(1)}_{p+1}(q)]\right]\,.
\end{eqnarray}

These expressions can be simplified by using standard recurrence relations valid for arbitrary Bessel functions ${\mathcal C}_{p}(q)$.
Indeed, using the short-hand notation $ {\mathcal C}_{(p,n)}(q) \equiv {\mathcal C}_{p-n}(q)+ {\mathcal C}_{p+n}(q)$,
${\mathcal C}_{[p,n]}(q) \equiv {\mathcal C}_{p-n}(q)- {\mathcal C}_{p+n}(q) $, we have
\beq
\label{sym_alt_exp}
\begin{array}{cll}
{\mathcal C}_{(p,1)}(q)&=&\frac{2p}{q}{\mathcal C}_{p}(q)\cr
{\mathcal C}_{[p,1]}(q)&=& 2{\mathcal C}'_{p}(q)\cr
{\mathcal C}_{(p,2)}(q)&=& 2\left(\frac{2p^2}{q^2}-1\right){\mathcal C}_p(q)-\frac{4}{q}{\mathcal C}_p'(q)\cr
{\mathcal C}_{[p,2]}(q)&=& \frac{4p}{q} \left[{\mathcal C}'_{p}(q)-\frac{1}{q}{\mathcal C}_{p}(q)\right]\cr
{\mathcal C}_{(p,3)}(q)&=& -{\mathcal C}_{(p,1)}(q)+\frac{2p}{q}{\mathcal C}_{(p,2)}(q)-\frac{4}{q}{\mathcal C}_{[p,2]}(q)\cr
{\mathcal C}_{[p,3]}(q)&=&  -{\mathcal C}_{[p,1]}(q)+\frac{2p}{q}{\mathcal C}_{[p,2]}(q)-\frac{4}{q}{\mathcal C}_{(p,2)}(q)\,.
\end{array}
\eeq
Using these relations, 
one can simplify the expressions \eqref{generali} above into
\begin{eqnarray}
\label{rel_x2y2xx_final2}
(x^2)_\omega &=&-\frac{2\pi}{u^2v_\infty^7}\left[ \left(-ue+\frac{u}{e}+i \right)H_{\frac{iu}{e}}^{(1)}(iu)\right. \nonumber\\
&&\left. +u (e^2-1)H_{\frac{iu}{e}+1}^{(1)}(iu)\right] \nonumber\\
(y^2)_\omega &=& -\frac{2\pi (e^2-1)}{e^2 v_\infty^7 u^2}\left[\left(\frac{u}{e}+i  \right) H_{\frac{iu}{e}}^{(1)}(iu) \right. \nonumber\\
&& \left. - u  H_{\frac{iu}{e}+1}^{(1)}(iu) \right]\nonumber\\
(x y)_\omega &=& \frac{2\pi \sqrt{e^2-1}}{u^2v_\infty^7}\left[ \left(-ue+\frac{u}{e}+i \right)H_{\frac{iu}{e}}^{(1)}(iu)\right. \nonumber\\
&& \left. -ie H_{\frac{iu}{e}+1}^{(1)}(iu)\right]\,,
\end{eqnarray}
where we introduced the  variable $u \equiv - iq=\frac{\omega}{\bar n} e $. 

Using the above expressions, one can, in principle, expand $\hat I_{ij}(\omega)$ in powers of $\e =1/e$.
In the text, we show the result so obtained at leading order in $\e$. The higher-order terms in $\e$ 
are more complicated and involve derivatives of the Hankel's functions {\it with respect to the order}
(in the vicinity of orders 0 or 1).

\section{Direct dynamical computation of $\chi_{\rm tail}$ by using ${\mathcal F}^i_{\rm (nonloc)}$} \label{RungeLenz}

In this Appendix we consider the traditional approach to computing the tail contribution to the 4PN scattering function; namely, we study the effect of adding to the Hamilton equations of motion derived from the  local Hamiltonian $H_{\rm loc}^{\rm 4PN}$, the additional {\it nonlocal} (tail) force ${\pmb{ \mathcal F}}^i_{\rm (nonloc)}$ , Eq. \eqref{F_non_loc}. As in the main text, this can be done simply by adding the effect of ${\pmb{ \mathcal F}}^i_{\rm (nonloc)}$ on the {\it Newtonian level} dynamics. The main conceptual difference with the time-localization
technique used in the main text is that, here, we shall time-localize tail effects at the level of the equations of motion, while, in Sec. \ref{sectail2}
we used a time-localized Hamiltonian tail action to compute $\chi_{\rm tail}$.

We then consider the following tail-perturbed Newtonian equations of motion
\beq
\frac{d {\mathbf p}}{dt}={\widehat {\mathbf F}}_{\rm (tot)}={\widehat {\mathbf F}}_{\rm (Newton)}+ {\widehat {\pmb{ \mathcal F}}}_{\rm nonloc}=-\frac{{\mathbf n}}{r^2}+{\widehat {\pmb{ \mathcal F}}}_{\rm nonloc}\,,
\eeq
where ${\mathbf n}=e_{\hat r}$, and where we used scaled variables ${\mathbf r}, {\mathbf p}$ and  $\mu$-rescaled forces: ${\widehat {\mathbf F}} \equiv  {\mathbf F}/\mu$, ${\widehat {\pmb{ \mathcal F}}}  \equiv {\pmb{ \mathcal F}}/\mu$.

The tail contribution to $\chi$ can then be computed by considering the evolution of the 
Laplace-Lagrange-Runge-Lenz, eccentricity vector (with ${\mathbf j}={\mathbf r}\times {\mathbf p}$),
\beq
{\mathbf A}={\mathbf p}\times {\mathbf j}-{\mathbf n}\,,
\eeq
We recall that, in absence of perturbations, ${\mathbf A}$ is conserved: its magnitude is equal to the eccentricity,
$e=\sqrt{1+2\bar Ej^2}$, and  ${\mathbf A}$ is directed from the origin towards the periastron (i.e., the point of closest approach in the case of hyperbolic motion).

The instantaneous eccentricity vector ${\mathbf A}(t)$ evolves under the effect of the additional force ${\pmb{ \mathcal F}}_{\rm nonloc}$:
\begin{eqnarray}
\label{dA_dt}
\frac{d  {\mathbf A}}{ dt}&=& \widehat {\pmb{ \mathcal F}}_{\rm nonloc}\times {\mathbf j}+{\mathbf p}\times \left({\mathbf r}\times \widehat {\pmb{ \mathcal F}}_{\rm nonloc}\right)\\
&=& \widehat {\pmb{ \mathcal F}}_{\rm nonloc}\times {\mathbf j}+   {\mathbf r}  ({\mathbf p} \cdot \widehat {\pmb{ \mathcal F}}_{\rm nonloc}) -
 \widehat {\pmb{ \mathcal F}}_{\rm nonloc} ({\mathbf r}\cdot {\mathbf p})\,.\nonumber
\end{eqnarray}

We  specialize Eq. \eqref{dA_dt} to equatorial motion, i.e., ${\mathbf j}=j e_{\hat z}$.  With respect to an adapted Cartesian system 
 with the $x$ direction along the unperturbed apsidal line ($e_{\hat x}$ a unit vector directed from the origin to periastron). 
We find
$A_x=jp_y-x/r$, $A_y=-jp_x-y/r$ and
\begin{eqnarray}
\label{cart_eqns}
\frac{d A_x }{ dt}&=& j \widehat {\mathcal F}_{\rm (nonloc)}{}^y +\left(x \widehat  {\mathcal F}_{\rm (nonloc)}^y -y \widehat  {\mathcal F}_{\rm (nonloc)}^x\right) p_y\nonumber\\
\frac{d A_y }{ dt}&=& -j \widehat {\mathcal F}_{\rm (nonloc)}{}^x -\left(x \widehat  {\mathcal F}_{\rm (nonloc)}^y -y \widehat  {\mathcal F}_{\rm (nonloc)}^x\right) p_x\,,\nonumber\\
\end{eqnarray}
as well as 
\begin{eqnarray}
{\mathbf n} &=& \cos \phi e_{\hat x}+\sin \phi e_{\hat y} \nonumber\\
{\mathbf p} &=& \frac{1}{j}[-\sin \phi e_{\hat x}+(\cos \phi +e)e_{\hat y}]\nonumber\\
p_r &=& ({\mathbf p}\cdot {\mathbf n})=\frac{e}{j}\sin \phi\,.
\end{eqnarray}
The squared magnitude of the perturbed eccentricity vector ${\mathbf A}(t)$ is  given by
\beq
A^2(t)=p^2j^2 +1 -\frac{2}{r}j^2\equiv 1+2\bar E (t) j(t)^2 \equiv e^2(t)\,,
\eeq
where $\bar E (t)= \frac{p^2}{2}-\frac{1}{r}$ 
denotes the perturbed energy ($j(t)$ being the perturbed angular momentum), and where we defined the perturbed eccentricity
as $ e(t)\equiv A(t) =\sqrt{1+2\bar E (t) j(t)^2}$.

As we are considering the {\it conservative} 4PN dynamics, both $\bar E(t)$ and $j(t)$ (and therefore $e(t)$) will be globally conserved.  I.e., $E_-=\bar E(t=-\infty)=\bar E(t=+\infty)=E_+$, and  $j_-=j(t=-\infty)=j(t=+\infty)=j_+$, as is easily checked  by using suitable integration by parts from the explicit expression of ${\pmb{ \mathcal F}}_{\rm nonloc}$. Therefore $e_-=e(t=-\infty)=e(t=+\infty)=e_+$.

While the magnitude of the perturbed eccentricity vector ${\mathbf A}(t)$ is globally conserved, its direction changes between $t=-\infty$ and $t=+\infty$ and this additional rotation of ${\mathbf A} $ encodes the tail contribution to the scattering function $\chi$.

The simplest way to see this fact, and to extract $\chi^{\rm tail}$ is , following section V D in Ref. \cite{Bini:2012ji}, 
to replace each vector of the equatorial  $(x,y)$ plane ${\mathbf V}=V_x e_{\hat x}+V_y e_{\hat y}$ by the corresponding complex number
$V=V_x+i V_y$.
The asymptotic values (at $t=\pm \infty$) of the complex numbers corresponding to ${\mathbf n}={\mathbf r}/r$ and ${\mathbf A}$ are easily found to satisfy
\begin{eqnarray}
n_\pm &=& e^{i\phi_\pm} \,, \nonumber\\ 
A_- &=& (-1+ipj)n_-\,,\quad A_+ =-(1+ipj)n_+\,,
\end{eqnarray}
where $\phi_\pm$ are the asymptotic values of the polar angle and where we have denoted for brevity as $p=p_{\pm \infty}= \sqrt{2 \bar E_{\pm}}$ and $j=j_\pm$ their asymptotic values.
Taking the ratio ${A_+}/{A_-}$ then yields
\begin{eqnarray}
\frac{A_+}{A_-}&=&\frac{-(1+ipj)e^{i\phi_+}}{(-1+ipj)e^{i\phi_-}}=\frac{e^{i{\rm arctan}(pj)}e^{i\phi_+}}{e^{-i{\rm arctan}(pj)} e^{i\phi_-}}\nonumber\\
&=&
e^{i(\phi_+-\phi_- + 2 {\rm arctan}(pj) )}\nonumber\\
&=&e^{i[\phi_+-\phi_- -\pi +2\left(\pi/2 + {\rm arctan}(pj)\right) ]}\,.
\end{eqnarray}
This can be rewritten as
\beq
\frac{A_+}{A_-}=e^{i(\chi-\chi^{\rm N}(\bar E,j))}
\eeq
where $\chi\equiv \phi_+-\phi_- -\pi$ is by definition the total (perturbed) scattering angle and where
\beq
\chi^{\rm N}(\bar E,j)\equiv{\rm arctan}\frac{1}{p j}\equiv{\rm arctan}\frac{1}{\sqrt{2\bar E_\pm j_\pm^2}}\,.
\eeq
As we see, the total angle of rotation of the ${\mathbf A}$ vector from $t=-\infty$ to $t=+\infty$, say $\alpha$, is equal to
\beq
\label{chi_pert}
\alpha=\chi^{\rm pert}(\bar E,j)-\chi^{\rm N}(\bar E,j)\equiv \chi^{\rm tail}(\bar E,j)\,,
\eeq
where the last equality is just the usual {\it definition} of the tail contribution to the scattering function $\chi(\bar E,j)$
(within our present context where $\chi^{\rm loc}(\bar E,j) = \chi^{\rm N}(\bar E,j)$).
Neglecting second order effects in $\widehat {\pmb{ \mathcal F}}_{\rm nonloc}$ in Eqs. \eqref{cart_eqns} and  \eqref{chi_pert},
we finally get
\begin{eqnarray}
e\chi^{\rm tail}(\bar E, j)&=& \int_{-\infty}^\infty \frac{d  A_y}{ dt}  dt \nonumber\\
&=&
\int_{-\infty}^\infty dt \left[(y(t)p_x(t)-j)\widehat {\mathcal F}^x_{\rm (nonloc)}\right. \nonumber\\
&& \left. -x(t)p_x(t) \widehat {\mathcal F}^y_{\rm (nonloc)}\right]\,.
\end{eqnarray}
To first order in $\widehat {\pmb{ \mathcal F}}_{\rm nonloc}$, we can evaluate the right-hand side of this formula along the unperturbed
(Newtonian) motion. Let us see how this allows one to compute the large-eccentricity (or, equivalently,
large-$j$) expansion of $\chi^{\rm tail}(\bar E, j)$.

Recalling the standard polar representation of
the Newtonian motion (in rescaled variables)
\beq
r(t)=\frac{j^2}{1+e\cos(\phi)}  \,,\qquad \frac{d\phi}{dt}=\frac{1}{j^3}(1+e\cos\phi)^2\,,
\eeq 
and the explicit expression of $\widehat {\mathcal F}_{\rm (nonloc)}{}_i $ \cite{Damour:2014jta}
\beq
\widehat {\mathcal F}_{\rm (nonloc)}{}_i =-\frac45 r(t) {\rm Pf}_{2s/c}\int\frac{d\tau}{|\tau|} n^i(t)I_{ij}^{(6)}(t+\tau)
\eeq
we find 
\beq
\label{dA_dt_finale}
e\chi^{\rm tail}(\bar E, j)=  
\oint dt {\rm Pf}_{2s/c}\int\frac{d\tau}{|\tau|}\,\, \delta(t,\tau) 
\eeq
where  
\beq
\label{dA_dt_finale1}
\delta(t,\tau)= \alpha^x (t)\, n^j(t)I_{xj}^{(6)}(t+\tau)  +\alpha^y(t)\, n^j(t)I_{yj}^{(6)}(t+\tau)  \,,
\eeq
with
\begin{eqnarray}
\label{dA_dt_finale2}
\alpha^x(t)&=& -\frac45 r  (y  p_x -j)\nonumber\\
&=& -\frac{4j^3}{5 (1+e\cos \phi  )^2}(\cos^2 \phi-e\cos \phi -2)\nonumber\\
\alpha^y(t)&=& \frac45 r  x p_x \nonumber\\
&=&-\frac{4j^3}{5(1+e\cos \phi )^2} \sin \phi\cos \phi \,,
\end{eqnarray}
where each orbital function here (apart from $j$ which is a constant of the motion) depends on $t$, namely, $r=r(t)$, $\phi=\phi(t)$, $x=x(t)$, $y=y(t)$, $p_x=p_x(t)$, $p_y=p_y(t)$.

By expanding the above quantities in inverse powers of $e$ (or, equivalently, remembering $e=\sqrt{1+2\bar E j^2}$,
in inverse powers of $j$, at fixed energy), we can, in principle, compute the  nonlocal contribution $\chi^{\rm tail}(\bar E, j)$ to 
any order in $1/e$. Here, we will do this calculation at the leading order in $1/e$.

For large $j$ (at fixed $\bar E$, so that $e=\sqrt{1+2\bar E j^2}\approx \sqrt{2\bar E} \, j$), we use the inverse eccentricity expansion
of the angular motion $\phi(t)$:  
\begin{eqnarray}
\label{hyp_eqs}
\phi(t)&=&\arctan(\tilde n t)+\frac{1}{e} \left[\frac{\tilde n t}{ \sqrt{1+\tilde n^2 t^2}}\right.\nonumber\\ 
&& \left.+\frac{{\rm arcsinh}(\tilde n t))}{1+\tilde n^2 t^2}\right]+O\left(\frac{1}{e^2}\right)\,,
\end{eqnarray}
with $\phi(0)=0$ and $r(0)=r_{\rm (peri)}$ and
\beq
 \tilde n\equiv \frac{(2\bar E)^{3/2}}{\sqrt{1+2\bar E j^2}}\approx \frac{2\bar E}{j}\,.
\eeq
Rescaling  the temporal variables as $\hat t=\tilde n \, t$ and $\hat \tau =\tilde n \tau$, 
a straightforward calculation (using Eqs. \eqref{hyp_eqs} in \eqref{dA_dt_finale}, \eqref{dA_dt_finale1} and \eqref{dA_dt_finale2}), shows that the crucial integrand $\delta(t,\tau)$ giving the value of $e\chi^{\rm tail}(\bar E, j)$ has a large-eccentricity expansion that starts as
\beq \label{expdelta}
\delta(\hat t,\hat \tau)= \frac{\nu}{j^{11}} \left[ e^8 f_8(\hat t,\hat \tau) + e^7 f_7(\hat t,\hat \tau) + O(e^6) \right],
\eeq
where
\begin{eqnarray}
   f_8(\hat t,\hat \tau)&=& -  \frac{544 }{5 [1+(\hat t+\hat \tau)^2]^{11/2}}\left[
\frac{12}{17}+\left(\hat t +\frac{2}{17}\hat \tau  \right)(\hat t +\hat \tau)^5\right. \nonumber\\
&&  -\frac{271}{68}\left(\hat t -\frac{344}{271}\hat \tau  \right)(\hat t +\hat \tau)^3   \nonumber\\
&& \left. 
-\frac{291}{68}\left(\hat t +\frac{187}{97}\hat \tau  \right)(\hat t +\hat \tau)
\right]
\end{eqnarray}
and $f_7= f_7^{\rm no-logs} + f_7^{\rm log}$, where
\beq
f_7^{\rm no-logs}= \frac{P_1}{\sqrt{1+\hat t^2}(1+(\hat t+\hat \tau)^2)^{11/2}}+\frac{P_2}{(1+(\hat t+\hat \tau)^2)^{6}} \,,
\eeq
and
\begin{eqnarray}
f_7^{\rm log} &=&   \frac{Q_1}{(1+(\hat t+\hat \tau)^2)^{13/2}} \ln(\hat t+\hat \tau+\sqrt{1+(\hat t+\hat \tau)^2})\nonumber\\
&&+ \frac{Q_2}{(1+(\hat t+\hat \tau)^2)^{11/2}}\ln (\hat t+\sqrt{1+\hat t^2})\,.
\end{eqnarray}
The coefficients entering the non-logarithmic and logarithmic parts of $f_7$ are
\begin{widetext}
\begin{eqnarray}
P_1&=& \left(-\frac{128}{5} \hat t^2+\frac{64}{5}\right)\hat \tau^6+\left(-\frac{1248}{5} \hat t^3+\frac{864}{5} \hat t\right)\hat \tau^5
+\left(\frac{2752}{5}+\frac{736}{5} \hat t^2-864 \hat t^4\right)\tau^4\nonumber\\
&+& \left(\frac{504}{5} \hat t^3+\frac{6088}{5} \hat t-1472 \hat t^5\right)\hat \tau^3+\left(-1344 \hat t^6-\frac{4488}{5}+\frac{4776}{5} \hat t^4+\frac{6408}{5}\hat t^2\right)\hat \tau^2\nonumber\\
&+& \left(-\frac{3168}{5}\hat t^7-\frac{6816}{5} \hat t+680 \hat t^3+\frac{7048}{5} \hat t^5\right)\hat \tau+\frac{384}{5}-\frac{608}{5} \hat t^8+568 \hat t^6+\frac{328}{5} \hat t^4-\frac{2736}{5} \hat t^2\nonumber\\
P_2&=& -\frac{192}{5} \hat t\hat \tau^7+\left(-\frac{1344}{5} \hat t^2+\frac{688}{15}\right)\hat \tau^6+\left(-\frac{8272}{5} \hat t-\frac{4032}{5} \hat t^3\right)\hat \tau^5+\left(-\frac{35544}{5}-8960 \hat t^2-1344 \hat t^4\right)\tau^4\nonumber\\
&+& \left(-\frac{55136}{3} \hat t^3-\frac{87816}{5} \hat t-1344 \hat t^5\right)\hat \tau^3+\left(\frac{44144}{5}-18608 \hat t^4-\frac{4032}{5}\hat t^6-\frac{50184}{5} \hat t^2\right)\tau^2\nonumber\\
&+& \left(-\frac{1344}{5} \hat t^7-\frac{46864}{5}\hat t^5+\frac{20904}{5}\hat t^3+\frac{68944}{5} \hat t\right)\hat \tau\nonumber\\
&-& \frac{28256}{15} \hat t^6+4960 \hat t^2-\frac{9728}{15}+\frac{18816}{5} \hat t^4-\frac{192}{5} \hat t^8 \nonumber \\
\end{eqnarray} 
\begin{eqnarray}
Q_1 &=& -\frac{608}{5} \left[-\frac{100}{19} \left(\hat t+\frac{1}{10}\hat \tau  \right) \left(\hat t+\hat \tau\right)^6+\frac{715}{19} \left(\hat t+\hat \tau\right)^4 \left(\hat t-\frac{118}{143}\hat \tau \right)\right.\nonumber\\
&& \left. +\frac{470}{19} \left(\hat t+\frac{1285}{376}\hat \tau  \right) \left(\hat t+\hat \tau\right)^2-\frac{825}{38}\hat \tau-\frac{345}{19} \hat t\right] 
\nonumber\\
Q_2 &=& -96 \left[\frac{9}{2}+\left(\hat t+\hat \tau \right)^4-\frac{41}{4} \left(\hat t+\hat \tau\right)^2\right] (\hat t+\hat \tau)\,.
\end{eqnarray}

By first evaluating the partie finie in $\hat \tau$ (with the  rescaled regularization scale  $2s\tilde n/c$) we find the intermediate result

\begin{eqnarray}
 {\rm Pf}_{2s \tilde n/c}\int_{-\infty}^\infty\frac{d \hat \tau}{|\hat \tau|}\,f_8(\hat t,\hat \tau) &=&  \frac{96}{5(1+\hat t^2)^{9/2}}\left[\left(-\frac{34}{3}\hat t^4+\frac{113}{2}\hat t^2-8\right)\ln[2(1+\hat t^2)]
-\left(-\frac{68}{3}\hat t^4-16+113 \hat t^2\right) \ln(2s\tilde n)\right. \nonumber\\
&&\left.  
+\hat t^6+\frac{1021}{18}\hat t^4-\frac{1159}{6}\hat t^2+\frac{137}{6}\right] \,.
\end{eqnarray}

\end{widetext}

But then, the explicit evaluation of the remaining $\hat t$ integral yields a vanishing contribution at order $e^8$:
\beq
\int_{-\infty}^\infty \frac{d \hat t}{\tilde n} \,  {\rm Pf}_{2s \tilde n/c}\int_{-\infty}^\infty \frac{d\tau}{|\tau|}\,f_8(\hat t,\hat \tau) =0\,.
\eeq
The first non-zero contribution to $ e  \chi^{\rm tail}$ comes from the $e^7 f_7$ term in $\delta(\hat t, \hat \tau)$, Eq. \eqref{expdelta}.
We managed to compute the double $\hat t$ and $\hat \tau$ integral of $ f_7(\hat t, \hat \tau)$ by first integrating by parts 
with respect to $\hat t$ (keeping $\hat \tau$ fixed) the logarithmic
terms in $f_7^{\rm log}$, i.e. by rewriting  $f_7^{\rm log}(\hat t, \hat \tau) = f_7^{\rm log-new}(\hat t, \hat \tau) + \frac{d}{d \hat t} g_7(\hat t, \hat \tau)$ where
\begin{eqnarray}
f_7^{\rm log-new}&=& -H_1(\hat t,\hat \tau)\frac{d}{d\hat t}\ln(\hat t+\hat \tau+\sqrt{1+(\hat t+\hat \tau)^2})\nonumber\\
&& -H_2(\hat t,\hat \tau)\frac{d}{d\hat t}
\ln (\hat t+\sqrt{1+\hat t^2})
\end{eqnarray}
with $H_1=\int^{\hat t} h_1 d\hat t$, etc. Because of the final $\hat t$ integration, the term $g_7(\hat t, \hat \tau)$ is checked to yield
a vanishing boundary contribution. After this transformation, we could perform all the needed integrations, with the final result
(using $e \approx \sqrt{2\bar E} j=v_\infty j $ and $\tilde n \approx \frac{2\bar E}{j}=\frac{v_\infty^2}{j}$)
\begin{eqnarray}
 \chi^{\rm tail} &=& -\frac{2}{5}\pi \frac{v_\infty^4}{j^4}\left[\frac{337}{3}+37\ln \left( \frac{s v_\infty^2}{4j c}\right)  \right]\nonumber\\
&+& O\left(\frac{v_\infty^3}{j^5}\right) \,.
\end{eqnarray}
The agreement of this purely dynamical evaluation of $ \chi^{\rm tail}$ with our previous result, Eq. \eqref{chitailLO}, obtained via
our time-localization of $ H^{\rm tail}$, is an additional check of the general time-localization technique introduced in \cite{Damour:2015isa} 
for bound motions, and extended in the present work to unbound motions.

\section{An alternative method to compute the local part of the scattering angle} \label{alternativemethod}

We sketch here the alternative method for computing the PN expansion of the polar equation $r=r(\phi)$ of the motion
introduced in Ref. \cite{Bini:2012ji}, and we show how we used it to check our computation of the contribution of
the {\it local} Hamiltonian $H^{\rm loc}$ to scattering.

By squaring Eq. \eqref{U_def}, we get the following differential equation  for the polar equation $\phi = \phi(u)$
of the orbit (here $u\equiv 1/r$) 
\beq
\label{eq_phi_u}
\left( \frac{d\phi}{du} \right)^2=\frac{j^2}{R''(\bar E,j,u)}\,,
\eeq
where the function $R''(\bar E,j,u)$ is obtained from the function $R(\bar E,j,u)$ describing 
the squared radial momentum,
\beq
p_r^2=R(\bar E,j,u)\,,
\eeq
via the definitions
\begin{eqnarray}
R'(\bar E, j,u)&=&\frac{1}{u^2}\frac{\partial }{\partial j^2} R(\bar E,j,u)\nonumber\\
R''(\bar E,j,u)&=&\frac{R(\bar E,j,u)}{\left(R'(\bar E,j,u)\right)^2}\,.
\end{eqnarray}

In the Newtonian approximation, the function $R''(\bar E,j,u)$ is a quadratic polynomial in $u$, namely
\beq
R''_{\rm Newton}(\bar E,j,u)= 2 \bar E + 2 u - j^2 u^2,
\eeq
 so that 
Eq. \eqref{eq_phi_u} can be easily integrated. The basic idea of the alternative approach to scattering
discussed in this Appendix (and introduced in Ref. \cite{Bini:2012ji}) is to transform the polar equation Eq. \eqref{eq_phi_u}
into a Newtonianlike equation by means of an appropriate change of variable $u \to \bar u$.

Under a change of  variable  $u=F(\bar u) = \bar u + O(\bar u^2)$,  Eq. \eqref{eq_phi_u} gets transformed  into  
\beq \label{polareq}
\frac{d\phi^2}{j^2}= \frac{du^2}{R''(u)}= \frac{d \bar u^2}{\bar R''(\bar u)}\,,
\eeq
where
\beq
\bar R''(\bar u)= R''(F(\bar u)) \left(\frac{d\bar u}{du}\right)^2\,.
\eeq
The idea is then to define the transformation $u=F(\bar u)$ such that  $\bar R''$ is reduced to a quadratic function of $\bar u$:
\beq \label{quad}
\bar R''(\bar u)=A+2B\bar u +C \bar u^2 \,
\eeq
with $A= 2\bar E+ O(\eta^2)$, $B=1+ O(\eta^2)$, and $C=-j^2 + O(\eta^2)$.
In view of Eq. \eqref{polareq}, the requirement \eqref{quad} yields the following condition on the transformation $u\to \bar u$:
\beq \label{condition}
(A+2B\bar u +C \bar u^2 )\left( \frac{du}{d\bar u} \right)^2=R''|_{u=u(\bar u)}\,.
\eeq
As emphasized in \cite{Bini:2012ji}, the three coefficients $A,B,C$ (all functions of $\bar E$, $j$ and $\nu$) entering the reduced quadratic polar equation \eqref{polareq}
contain all the information needed to compute gauge-invariant measures of the relativistic orbit, such as periastron advance (in the bound case)
and the scattering angle (in the unbound case).  Indeed, the solution of
\beq
\left(\frac{d\phi}{d\bar u} \right)^2= \frac{j^2}{(A + 2B \bar u + C  \bar u^2)}
\eeq
is the following precessing conic
\beq 
\bar u= \frac1{\bar r}= u_p \left(1+ \bar e \cos \frac{\phi}{K}\right)  \,,\qquad u_p \equiv\frac{1}{p}\,,
\eeq
with
\beq
A= \frac{u_p^2j^2(\bar e^2-1)}{K^2}\,,\qquad B=\frac{u_p j^2}{K^2}\,,\qquad C=-\frac{j^2}{K^2} \,.
\eeq
The above relations are easily inverted leading to
\beq
K^2=-\frac{j^2}{C}\,,\qquad u_p=-\frac{B}{C} \,,\qquad \bar e^2-1=-\frac{AC}{B^2}\,.
\eeq
The periastron-to-periastron precession angle is $\Phi= 2\pi K$, while the scattering angle $\chi$ is given by
\beq
\chi+\pi =2K \arccos\left( -\frac{1}{\bar e} \right)\,.
\eeq
Note that the function $K(\bar E,j)$ enters both $\Phi$ and $\chi$, though one has to analytically continue it from $\bar E<0$
to $\bar E>0$.

By considering the large-separation limit ($u \to \infty$) of the EOB dynamics, one finds that one has simply
\beq
A=2\bar E \,.
\eeq
The PN expansion  of the transformation $u=F(\bar u)$ is found to have the form
\begin{widetext}
\begin{eqnarray}
u &=& \bar u- \bar u^2\eta^2+\left[ \left(\frac{3}{2}\nu+\frac{3}{4}\right)\bar u^3+\left(\frac{5}{2}\nu-\frac{17}{4}\right)\frac{\bar u^2}{j^2}  \right]\eta^4\nonumber\\
&+& \left\{ 
\left(-3\nu+\frac{9}{4}\nu^2\right) j^2\bar u^5+\left(-\frac{25}{4}\nu^2+11\nu-\frac{1}{2}\right)\bar u^4\right. \nonumber\\
&& \left. +\left[\left(-\frac{27}{4}\nu^2+9\nu\right)\bar E+\left(\frac{13}{2}-\frac{7}{8}\nu^2+\left(-\frac{41}{128}\pi^2+\frac{25}{3}\right)\nu\right)\frac{1}{j^2}\right]\bar u^3\right. \nonumber\\
&+& \left.
\left[\left(-\frac{23}{4}\nu^2+27\nu-9\right)\frac{\bar E}{j^2} +\left(-\frac{71}{2}-\frac{35}{8}\nu^2+\left(-\frac{205}{128}\pi^2+\frac{245}{3}\right)\nu\right)\frac{1}{j^4}\right]\bar u^2
\right\}\eta^6\nonumber\\
&+&  f (\bar u) \sqrt{2\bar E-j^2\bar u^+2\bar u}\, \eta^8
\end{eqnarray}
where we have left unspecified  at $O(\eta^8)$ the 4PN-level transformation  $f(\bar u)$ for reasons which will become clear soon.
The corresponding coefficients $B$ and $C$,  are found to have the form
\begin{eqnarray}
B &=& 1+4\bar E\eta^2+\eta^4 (17-10\nu)\frac{\bar E}{j^2} +\eta^6\left[(23\nu^2-108\nu+36)\frac{\bar E^2}{j^2} +\left(\frac{205}{32}\nu\pi^2-\frac{980}{3}\nu+\frac{35}{2}\nu^2+142\right)\frac{\bar E}{j^4}\right]+B_8\eta^8\nonumber\\
C &=& -j^2+6\eta^2+\eta^4 \left[(15-6\nu)\bar E+\left(\frac{51}{2}-15\nu\right)\frac{1}{j^2}\right]\nonumber\\
&+& \eta^6\left\{(-12\nu+9\nu^2)\bar E^2+\left[ 180+45\nu^2+\left(\frac{123}{32}\pi^2-382\right)\nu\right]\frac{\bar E}{j^2} +\left[213+\frac{105}{4}\nu^2+\left(-490+\frac{615}{64}\pi^2\right)\nu\right]\frac{1}{j^4}\right\}\nonumber\\
&+& C_8\eta^8\,,
\end{eqnarray}
where we left unspecified the 4PN-level coefficients $B_8$ and $C_8$.

The formal solution of the condition  \eqref{condition} for the 4PN-level function $f (\bar u)$ can be cast in the form
\beq \label{solf}
f (\bar u)= \int_0^{\bar u} \frac{\sigma( u)}{Q( u)^{3/2}}\, d u\,,\qquad\qquad Q(\bar u)=2\bar E-j^2\bar u^2+2\bar u=j^2(u_{\rm (max)}-\bar u)(\bar u-u_{\rm (min)})\,.
\eeq
Here $\sigma (u)$ is a 4PN-level ``source term" coming from $R''(u)$ (itself derived from the 4PN-accurate local Hamiltonian). It is such that $\sigma(0)=0$, and it comprises both a log-part, $\sigma^{\ln{}}(u)$ (involving $\ln u$), and a log-free part, $\sigma^{\rm no-log}(u)$.
The latter log-free part reads
\beq
\sigma^{\ln{}}(u)=\nu \frac{16}{15}u^4\left(-62 u +37 j^2 u^2 -74 \bar E\right)\ln (u s)\,.
\eeq
while $\sigma^{\rm no-log}(u)$ is an 8th degree polynomial (with coefficients depending on $\bar E$, $j$ and $\nu$), having a simple root at $u=0$. Its explicit value reads
\begin{eqnarray}
\sigma^{\rm no-log}(u)&=& \left(\frac{9}{2}\nu+\frac{27}{2}\nu^2-15\nu^3\right) j^6 u^8+\left(105\nu^3-27\nu-183\nu^2\right) j^4 u^7\nonumber\\
&+&\left\{(-27\nu+90\nu^3-81\nu^2) j^4\bar E+\left[-\frac{5}{4}-240\nu^3+\left(\frac{1453}{2}-\frac{123}{32}\pi^2\right)\nu^2+\left(\frac{167}{90}+\frac{25729}{3072}\pi^2\right)\nu\right] j^2\right\} u^6\nonumber\\
&+& \left[\left(\frac{1581}{2}\nu^2+30\nu-420\nu^3\right) j^2\bar E+\frac{45}{2}+180\nu^3+\left(\frac{451}{64}\pi^2-\frac{5153}{6}\right)\nu^2+\left(\frac{14383}{72}-\frac{60251}{3072}\pi^2\right)\nu\right] u^5\nonumber\\
&+& \left\{(-180\nu^3+162\nu^2+54\nu) j^2\bar E^2\right. \nonumber\\
&+& \left. \left[-\frac{397}{16}+480\nu^3+\left(-\frac{6451}{4}+\frac{123}{16}\pi^2\right)\nu^2+\left(-\frac{23761}{1536}\pi^2+\frac{39547}{180}\right)\nu\right]\bar E\right. \nonumber\\
&+& \left.\left[-\frac{5889}{32}+\frac{597}{8}\nu^2+\left(\frac{123}{64}\pi^2-\frac{661}{8}\right)\nu\right]\frac{1}{j^2}\right\} u^4\nonumber\\
&+& \left\{(420\nu^3-867\nu^2+72\nu)\bar E^2+\left[\frac{813}{4}+\frac{457}{2}\nu^2+\left(\frac{369}{32}\pi^2-1083\right)\nu\right]\frac{\bar E}{j^2}\right.\nonumber\\ 
&+&\left.\left[\frac{1425}{2}+120\nu^2+\left(-\frac{4430}{3}+\frac{205}{8}\pi^2\right)\nu\right]\frac{1}{j^4}\right\} u^3\nonumber\\
&+& \left\{(-36\nu-108\nu^2+120\nu^3)\bar E^3+(-648\nu+216+138\nu^2)\frac{\bar E^2}{j^2}\right. \nonumber\\
&+& \left. \left[\frac{4275}{4}+180\nu^2+\left(\frac{615}{16}\pi^2-2215\right)\nu\right]\frac{\bar E}{j^4} -\frac12  C_8\right\} u^2-u B_8
\end{eqnarray}
Note that the so-far unspecified coefficients  $B_8$, and $C_8$ contribute the terms $ - B_8 u -\frac12  C_8 u^2$ to
$\sigma^{\rm non-log}(u)$.  Note also that the integral \eqref{solf} defining the function $f(u)$ entering the 4PN-level 
transformation $u=F(\bar u)$ is {\it a priori} singular at the two roots of $Q(\bar u)$, i.e., at the Newtonian 
(periastron and apastron) values of $u_{\rm (max/min)}$,
\beq
u_{\rm max}=\frac{1}{j^2}\left(1+\sqrt{1+2\bar E j^2}  \right)\,,\qquad u_{\rm min}=\frac{1}{j^2}\left(1-\sqrt{1+2\bar E j^2}  \right)\,.
\eeq
One finds that one must require the regularity of the transformation $u=F(\bar u)$ around these roots, and that this 
 regularity condition implies  the vanishing of the two following Hadamard-partie-finie integrals
 \beq
\label{regul_conds}
{\rm Pf} \int_0^{u_{\rm (max)}} \frac{\sigma( u)}{Q( u)^{3/2}} du=0\, ,\, {\rm Pf} \int_0^{u_{\rm (max)}} \frac{\sigma( u)}{Q( u)^{3/2}} du=0 \,.
\eeq
These conditions can be simplified by using the  identity
\beq
\frac{1}{Q^{3/2}}=\frac{1}{(1+2\bar E j^2)}\frac{d}{du}\left(\frac{uj^2-1}{Q^{1/2}}  \right)\equiv \frac{d}{du}g(u)\,,\qquad 
g(u)=\frac{1}{(1+2\bar E j^2)}\left(\frac{uj^2-1}{Q^{1/2}}  \right)
\eeq
to integrate by parts. This leads to the equivalent conditions
\beq
{\rm Pf} \int_0^{u_{\rm (max)}}g(u)\sigma'(u) du=0\,,\qquad {\rm Pf} \int_{u_{\rm (min)}}^{u_{\rm (max)}}g(u)\sigma'(u) du=0\,.
\eeq
These two (regularity) conditions completely determine the two parameters $B_8$ and $C_8$.
From the practical point of view, while the partie finie of the  integrals involving $\sigma^{\rm no-log}(u)$ can be straightforwardly computed, the partie finie of the  integrals involving $\sigma^{\ln{}}(u)$ can only be computed as an expansion in inverse powers of the eccentricity,
as was already the case for the corresponding  integrals $I_\chi$ and ${\mathcal I}_\chi$ of sec. III. More precisely, one finds, using the explicit expression,
\beq
\frac{d \sigma^{\ln{}}(u)}{du}=\frac{16}{15}\nu u^3 \left[2\ln (us)(111 u^2j^2-148 \bar E-155 u) +37 u^2j^2-74 \bar E-62 u \right]\,,
\eeq
 that the log-parts of the two above conditions, namely
 \beq
{\rm Pf} \int_0^{u_{\rm (max)}}g(u)\frac{d \sigma^{\ln{}}(u)}{du} du \,,\qquad {\rm Pf} \int_{u_{\rm (min)}}^{u_{\rm (max)}}g(u)\frac{d \sigma^{\ln{}}(u)}{du} du \,,
\eeq
form exactly the combinations $I_\chi$ and ${\mathcal I}_\chi$ of sec. III, Eqs. \eqref{I_chi_def} and \eqref{matcal_I_chi_def}, respectively.

Implementing the above method, we have computed $\chi^{\rm loc}$, and found that it  agreed with the result obtained by
the method explained in the main text. For concreteness, let us just indicate the beginnings of the PN expansions of $K$ and $\bar e$
(with the notation $\bar e_0(\bar E, j) \equiv\sqrt{1+2\bar E j^2}$)
\begin{eqnarray}
\label{bar_e_expr}
K&=& 1+\frac{3}{j^2}\eta^2+\left[\left(\frac{15}{2 } -  3\nu \right) \frac{\bar E}{j^2}+\left(\frac{105}{4 }-\frac{15}{2 }\nu\right) \frac{1}{j^4}\right]\eta^4\nonumber\\
&+& \left\{\left(-6\nu+\frac92\nu^2\right)\frac{\bar E^2}{j^2}+\left[\frac{45}{2}\nu^2+\left(\frac{123}{64}\pi^2-218\right)\nu+\frac{315}{2}\right]\frac{\bar E}{j^4}\right. \nonumber\\
&& \left. +\left[\frac{105}{8}\nu^2+\left(\frac{615}{128}\pi^2-\frac{625}{2}\right)\nu+\frac{1155}{4}\right]\frac{1}{j^6}\right\}\eta^6+O(\eta^8)\nonumber\\
\bar e &=&\bar e_0- \frac{2\bar E}{\bar e_0}
\left(4\bar E j^2+3  \right) \eta^2\nonumber\\
&& +\frac{\bar E}{\bar e_0^3}\left[ 64 \bar E^3 j^4+(52\nu-2)\bar E^2 j^2+(-70+56\nu)\bar E +\left(-\frac{51}{2}+15\nu\right)\frac{1}{j^2} \right]\eta^4 \nonumber\\
&& +\frac{\bar E}{\bar e_0^5}\left\{ -512\bar E^5 j^6+(176\nu+16-220\nu^2)\bar E^4 j^4+\left[-380-540\nu^2+\left(\frac{11104}{3}-\frac{533}{8}\pi^2\right)\nu\right]\bar E^3 j^2 \right.\nonumber\\
&& 
+\left[-1532-480\nu^2+\left(\frac{16612}{3}-\frac{1681}{16}\pi^2\right)\nu\right]\bar E^2+\left[-1061-185\nu^2+\left(\frac{8536}{3}-\frac{1763}{32}\pi^2\right)\nu\right]\frac{\bar E}{j^2}\nonumber\\
&& \left. +\left[-213-\frac{105}{4}\nu^2+\left(490-\frac{615}{64}\pi^2\right)\nu\right] \frac{1}{j^4}
 \right\} +O(\eta^8)  \,.
\end{eqnarray}

\end{widetext}

\section*{Acknowledgments}
DB thanks ICRANet and the italian INFN for partial support and IHES for warm hospitality at various stages during the development of the present project.

\end{document}